\documentclass[10pt,english,english,nofootinbib,notitlepage,superscriptaddress,aps,prd]{revtex4-1}

\usepackage{enumerate}
\usepackage{amsmath}
\usepackage{amsfonts}
\usepackage{amssymb}
\usepackage[utf8]{inputenc}
\usepackage[T1]{fontenc}
\usepackage{mathtools}
\usepackage{wasysym}
\usepackage{accents}
\usepackage{comment}
\usepackage[svgnames]{xcolor}
\usepackage[colorlinks=true, pdfstartview=FitV, linkcolor=blue, citecolor=red, urlcolor=magenta]{hyperref}
\usepackage{url}
\usepackage[customcolors]{hf-tikz}
\hfsetfillcolor{white}
\hfsetbordercolor{black}
\usepackage{graphicx}
\usepackage{tikz-cd}
\usepackage{subfig}
\usepackage{bm}
\usepackage{tikz}
\usetikzlibrary{arrows,decorations.markings}
\usepackage{framed}
\usepackage{xspace}
\usepackage[normalem]{ulem}
\newcommand\ie{\mbox{\textit{i.\,e.}}\xspace}
\newcommand\cf{\mbox{c.\,f.}\xspace}
\newcommand\eg{\mbox{e.\,g.}\xspace}
\newcommand\D{\mathrm{d}}

\DeclareMathOperator*{\SumInt}{%
\mathchoice%
  {\ooalign{$\displaystyle\sum$\cr\hidewidth$\displaystyle\int$\hidewidth\cr}}
  {\ooalign{\raisebox{.14\height}{\scalebox{.7}{$\textstyle\sum$}}\cr\hidewidth$\textstyle\int$\hidewidth\cr}}
  {\ooalign{\raisebox{.2\height}{\scalebox{.6}{$\scriptstyle\sum$}}\cr$\scriptstyle\int$\cr}}
  {\ooalign{\raisebox{.2\height}{\scalebox{.6}{$\scriptstyle\sum$}}\cr$\scriptstyle\int$\cr}}
}

\begin{document}
\newcommand{\Areia}{
\affiliation{Department of Chemistry and Physics, Federal University of Para\'iba, Rodovia BR 079 - Km 12, 58397-000 Areia-PB,  Brazil.}
}
\newcommand{\Lavras}{
\affiliation{Physics Department, Federal University of Lavras, Caixa Postal 3037, 37200-000 Lavras-MG, Brazil.}
}

\newcommand{\JP}{
\affiliation{Physics Department, Federal University of Para\'iba, Caixa Postal 5008, 58059-900, Jo\~ao Pessoa, PB, Brazil.}
}

\newcommand{\UOS}{
\affiliation{University of Szczecin, Department of Physics, Wielkopolska 15, PL-71-061 Szczecin, Poland.}
}

\newcommand{\US}{
\affiliation{Dipartimento di Ingegneria Industriale, Universit\`a degli Studi di Salerno, Via Giovanni Paolo II, 132 I-84084 Fisciano (SA), Italy.}
}

\newcommand{\INFN}{
\affiliation{INFN, Sezione di Napoli, Gruppo collegato di Salerno, Via Giovanni Paolo II, 132 I-84084 Fisciano (SA), Italy}
}

\title{Quantum-spacetime effects on nonrelativistic Schr\"odinger evolution}

\author{Fabian Wagner}\email{fwagner@unisa.it}
\UOS
\US
\INFN

\author{Gislaine Varão}\email{gislainevarao@gmail.com}
\JP

\author{Iarley P. Lobo}\email{lobofisica@gmail.com}
\Areia
\Lavras

\author{Valdir B. Bezerra}\email{valdir@fisica.ufpb.br}
\JP

\date{\today}

\begin{abstract}
The last three decades have witnessed the surge of quantum gravity phenomenology in the ultraviolet regime as exemplified by the Planck-scale accuracy of time-delay measurements from highly energetic astrophysical events. Yet, recent advances in precision measurements and control over quantum phenomena may usher in a new era of low-energy quantum gravity phenomenology. In this study, we investigate relativistic modified dispersion relations (MDRs) in curved spacetime and derive the corresponding nonrelativistic Schrödinger equation using two complementary approaches. First, we take the nonrelativistic limit, and canonically quantise the result. Second, we apply a WKB-like expansion to an MDR-inspired deformed relativistic wave equation. Both approaches imply equivalent results for single-particle quantum mechanics. Based on a map between our approach and the generalized uncertainty principle (GUP), we recognise in the latter the MDR which is least amenable to low-energy experiments. Consequently, importing data from time-delay measurements, we constrain the linear GUP up to the Planck scale and improve on current bounds to the quadratic one by 17 orders of magnitude. MDRs with larger implications in the infrared, however, can be tightly constrained in the nonrelativistic regime, from which we use the ensuing deviation from the equivalence principle to bound some MDRs to up to one order of magnitude below the Planck scale, while constraining those customarily associated with the bicrossproduct basis of the $\kappa$-Poincar\'e algebra to energy scales beyond $10^{15}$GeV.
\end{abstract}

\maketitle


\tableofcontents

\section{Introduction}

After more than a century of research, the quest for a theory of quantum gravity has led to the development of a wide-ranging and intricate landscape of ideas, as detailed in the recent comprehensive reviews \cite{Kiefer:2007ria,Loll:2022ibq}. Yet, despite the abundance of proposed approaches, none have provided compelling and decisive experimental evidence in their favour. This enduring lack of consensus can be attributed primarily to the dearth of experimental input due to the seeming inaccessibility of the pertinent length scale, the Planck length $\ell$. However, following advances of precision in observations and experiments, as well as growing proficiency in manipulating quantum phenomena \cite{Bose:2017nin,Marletto:2017kzi}, the realisation of a meaningful phenomenology of quantum gravity seems increasingly attainable \cite{Amelino-Camelia:1999hpv} (see \cite{Addazi:2021xuf} for an extensive review).

For example, in the near future we may have the ability to observe modifications to the kinematics of special relativity, implying MDRs.  The ensuing changes could entail violations of Lorentz invariance (LIV) which are summarised in the standard model extension \cite{Kostelecky:2008ts} or, somewhat less radically, deformations thereof known as doubly (or deformed) special relativity (DSR) \cite{Magueijo:2001cr,Amelino-Camelia:2000stu} (see \cite{Amelino-Camelia:2008aez,Arzano:2021scz,Kowalski-Glikman:2022xtr} for recent reviews). Even though the momentum space underlying these models is curved \cite{Kowalski-Glikman:2002oyi}, a recurring theme in quantum gravity phenomenology, they are inherently relativistic.

While quantum-gravity effects are famously important in regions of strong curvature, such as the beginning of the universe and the vicinity of the singularity inside black holes, this is just one of several possible amplifiers.\footnote{The importance of amplifiers for quantum gravity phenomenology was emphasised in \cite{Amelino-Camelia:2008aez}.} Quantum features of gravity can also be tested at the nonrelativistic level. This is particularly so for those MDRs with pronounced infrared (IR) effects \cite{Carmona:2000gd,Freidel:2021wpl} (see Sec. 5 of \cite{Amelino-Camelia:2008aez} for a review). In fact, some of the most promising soon-to-be feasible tests of perturbative quantum gravity \cite{Donoghue:1993eb,Donoghue:1994dn,Donoghue:1995cz} are of this kind \cite{Bose:2017nin,Marletto:2017kzi}. 

The most common framework for exploring nonrelativistic quantum-spacetime effects is through the use of quantum-mechanical minimal-length models \cite{Kempf:1994su,Maggiore:1993kv,Benczik:2002tt,Bosso:2020aqm,Petruzziello:2020wkd} (see \cite{Hossenfelder:2012jw,Tawfik:2014zca} and, most recently, chapter 3 of \cite{Wagner:2022bcy} for reviews, open problems have been surveyed in \cite{Bosso:2023aht}). In this context, Planck-scale effects are introduced into quantum mechanics by virtue of a minimal localisation of the position operator implied by a generalized uncertainty principle (GUP). Although the relationship between GUPs and curved momentum space has been studied recently \cite{Singh:2021iqa,Wagner:2021bqz,Wagner:2022rjg,Wagner:2022dkc} and their connection to MDRs has been previously emphasised \cite{Hossenfelder:2005ed}, a detailed derivation of the nonrelativistic quantum-equivalent of MDRs has yet to be established.

Furthermore, recent studies have demonstrated that spacetime-curvature induced effects of quantum gravitational origin may not necessarily suffer from the twofold suppression caused by the presence of the Planck scale and the weakness of the gravitational field, when other amplifiers are considered \cite{Amelino-Camelia:2020bvx}. For instance, models violating the weak equivalence principle may be tightly constrained due to the high experimental precision E\"otv\"os-like experiments can achieve nowadays \cite{MICROSCOPE:2022doy}. Consequences of spacetime curvature can be incorporated into the concept of deformed relativistic kinematics from the point of view of geometry by reverting to Finsler \cite{Girelli:2006fw,Gibbons:2007iu,Kostelecky:2011qz,Amelino-Camelia:2014rga} or Hamilton \cite{Barcaroli:2015xda} spaces (for a mathematical introduction see \cite{Miron:2002bns,Miron:2012na}, for reviews \cite{Pfeifer:2019wus,Albuquerque:2023icp}).

\begin{figure}[h!]
    \centering
    \includegraphics[width=.6\linewidth]{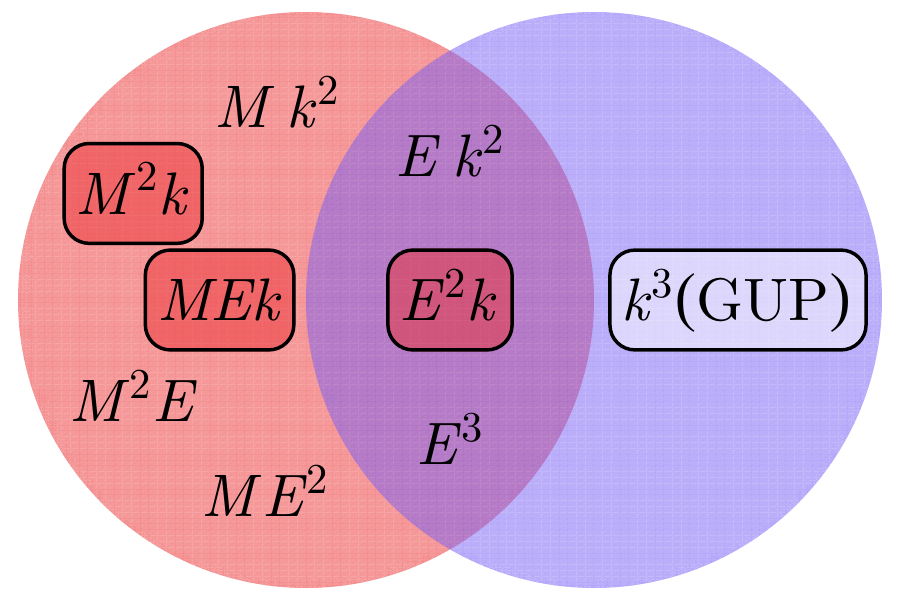}
    \caption{Dependence of correction terms to the relativistic dispersion relation on momentum $k,$ mass $M,$ and energy $E$ for all models at order $\ell.$ The violet (left) and red (right) circles indicate those which are amenable to relativistic (UV) and nonrelativistic (IR) measurements, respectively. Those which are particularly suitable for nonrelativistic reasoning are framed and have an additional red background. The GUP is indicated by a frame and with a white background.}
    \label{fig:UVIR}
\end{figure}

In this article, we derive deformations of the Schr\"odinger equation describing a nonrelativistic particle moving in curved (quantum-)spacetime from relativistic MDRs. Obtaining the correct nonrelativistic limit of a relativistic model is a nontrivial task, particularly when dealing with single-particle quantum dynamics. In the literature, there have been two main paths to deriving such a theory:
\begin{itemize}
    \item First, obtain the nonrelativistic Hamiltonian, then quantise it \`a la Dirac \cite{DeWitt:1952js,Wajima:1996cu,Tagirov:2002hw,Tagirov:2003kao,Petruzziello:2021vyf}. Dating back to 1952, this is the classic approach to the problem. However, while there is a suggestive kind of operator ordering, this physically relevant choice has to be made in a rather ad-hoc manner.
    \item Start with a relativistically covariant wave equation, afterwards derive the effective Schr\"odinger equation in a sort of WKB expansion \cite{Kiefer:1990pt,Tagirov:1990kao,Tagirov:1992kao,Lammerzahl:1995zz,Giulini:2012zu,Schwartz:2018pnh}. This corresponds to the single-particle sector of some underlying quantum field theory. Therefore, the ordering is fixed at the relativistic level, where it is based on clear physical principles.
\end{itemize} 
The above discussion is summarised by the following illustrative diagram 
\begin{equation}
\begin{tikzcd}
\text{dispersion relation} \arrow[d,"c\rightarrow \infty"'] \arrow[r,"\text{inspires}"]  & \text{wave equation} \arrow[d,"c\rightarrow \infty"']  \\
\text{nonrel. Hamiltonian} \arrow[r,"\text{quantise}"'] & \text{Schr\"odinger Eq.}
\end{tikzcd}\label{limitdiag}
\end{equation}
In this paper, we choose to follow both routes, and compare the results. 
Pursuing a model-agnostic approach, we consider all possible MDRs at first order in the additional length scale $\ell$ which contain positive powers of energy $E$, spatial momentum $k$ and mass $M,$ \ie
\begin{equation}
    M^2c^2=\left(\frac{E}{c}\right)^2-k^2+\ell [k^3],\label{eqn:introMDR}
\end{equation}
where $[k^3]$ stands for any mix of $Mc,$ $k$ and $E/c$ which has units of $k^3$ (the allowed combinations are displayed in Fig. \ref{fig:UVIR}). In so doing, we start applying a $3+1$-decomposition in ADM-variables \cite{Arnowitt:1959ah,Arnowitt:1960es,Arnowitt:1962hi} to the spacetime metric. For reasons of consistency, we assume the induced metric on the ensuing spacelike hypersurfaces to be time-independent. With this foliation in mind, we first take the nonrelativistic limit, in other words a $1/c$-expansion to order 0, without quantum-gravity corrections. As those corrections are suppressed by the length scale $\ell,$ they are added at leading order in $1/c$ in a second step.

We find that both approaches predict the same deformed Schr\"odinger equation. Given this effective nonrelativistic deformation, we rank the models by the amount to which they are amenable to experiments and observations in the IR. In Fig. \ref{fig:UVIR}, we display this ranking, further taking into account their suitability to observation in the ultraviolet (UV) regime. Remarkably, the MDR that introduces GUP-like corrections in the nonrelativistic limit ($k^3$) is well-suited for experimental testing only in the UV regime, rather than the IR. Nevertheless, there are MDRs which have wide implications at low energies, especially those which are linear in $k$. In other words, the one approach to nonrelativistic quantum gravity phenomenology which has been prevalent over the last three decades, the GUP, corresponds to exactly that MDR which is least suitable for nonrelativistic reasoning.

Delving further into the connection between the nonrelativistic limit of MDRs and minimal-length quantum mechanics, we uncover deformed Heisenberg algebras that correspond to the MDR models under investigation. Consequently, a more encompassing set of commutation relations emerges, surpassing the conventional considerations of minimal-length models, for instance, by introducing anisotropies. Remarkably, most of these commutation relations fail to recover the standard Heisenberg algebra as the momentum tends to zero (although they do so in the limit of vanishing $\ell$), presenting a harbinger of the UV/IR-mixing known from noncommutative geometries \cite{Szabo:2001kg,Minwalla:1999px}. This finding opens up new pathways towards quantum gravity phenomenology in the IR as indicated in \cite{Carmona:2000gd,Amelino-Camelia:2008aez,Freidel:2021wpl}. 

Inspired by the fact that conventional minimal-length models are those which are least amenable to low-energy experiments (\cf Fig. \ref{fig:UVIR}), we import bounds from UV observations, in particular, time delay measurements of light stemming from highly energetic astrophysical phenomena \cite{MAGIC:2017vah,MAGIC:2020egb}. Consequently, we constrain the GUP model parameters up to the Planck scale for the linear deformation \cite{Ali:2009zq,Ali:2011fa}, and improve on bounds to the quadratic one \cite{Kempf:1994su,Maggiore:1993kv} by 17 orders of magnitude. These results corroborate the expectation that the GUP is to be tested at high rather than low energies.

Finally, we illustrate the impact of MDRs on single-particle quantum dynamics by presenting several examples that highlight the phenomenological implications. First, we examine the Landau levels within arbitrary static background metrics; to provide an example, we consider the example of a cosmic string spacetime. We further derive the deformation of flat-space quantum mechanics, which holds relevance for table-top experiments. To present an example, we provide corrections to the spectrum of the harmonic oscillator, and estimate partially super-Planckian constraints on the MDR model parameters from an earlier experiment on the GUP \cite{Bawaj:2014cda}. Furthermore, an investigation of the weak equivalence principle in the context of the Schwarzschild background enables us to constrain certain MDRs to within one order of magnitude from the Planck scale, while bounds on those generally associated with the bicrossproduct basis of DSR \cite{Kowalski-Glikman:2002iba} reach up to four order of magnitude from it. These outcomes are testament to the power of nonrelativistic reasoning about Planck-scale effects. 

In short, we connect quantum gravity phenomenology for quantum dynamics in the nonrelativistic and relativistic regimes, thus interrelating the fields of MDRs (derived from DSR or LIV) and GUPs. As exemplified by the transfer of bounds and by applications, both the relativistic as well as the nonrelativistic sides can benefit significantly from this connection.

Throughout the paper, we set $\hbar=1$ unless otherwise stated, while, for obvious reasons, we retain the speed of light $c.$ Indices in Greek stand for spacetime coordinates, while Roman letters indicate spatial ones. The latter refer to a coordinate system which, as introduced in Sec. \ref{sec:cov}, is adapted to the nonrelativistic limit. Finally, we use the mostly positive metric signature. 


\section{Phenomenology of Modified Dispersion Relations}\label{sec:MDR}

Front and center in describing relativistic motion, be it in the context of special relativity or its deformed variants, are dispersion relations of the form
\begin{equation}
    \mathcal{C}(E,k,M)=M^2c^2,\label{GenDisRel}
\end{equation}
with the mass Casimir of the underlying spacetime symmetry algebra $\mathcal{C}.$ Throughout this paper, we will assume this Casimir to be of the form
\begin{equation}
    M^2c^2=\mathcal{C}_\ell=\mathcal{C}_0+\delta \mathcal{C},\label{eqn:mdr0}
\end{equation}
with the Casimir operator of the customary Lorentz algebra $\mathcal{C}_0,$ and its first-order deformation in $\ell,$ namely $\delta \mathcal{C},$ which, as in Eq. \eqref{eqn:introMDR}, is of the form $\ell [k^3].$ 

While a contribution to $\delta\mathcal{C}$ of the form $\ell M^3$ does not change the dispersion relation -- it amounts to a constant redefinition of the rest mass --, all other combinations of $M,$ $k$ and $E$ are, in principle, allowed. While nonanalytic corrections have been considered in the literature \cite{Freidel:2018apz}, for reasons of simplicity, we assume the dependence on the dynamical variables to be analytic. Under this assumption, the most general MDR at order $\ell$ reads
\begin{equation}
    M^2c^4=E^2-k^2c^2+\ell\sum_{n=0}^2\sum_{m=0}^{3-n} a_{n,m,3-n-m}(Mc)^n k^m \left(\frac{E}{c}\right)^{3-n-m},\label{mdr1}
\end{equation}
where the dimensionless constants $a_{n,m,3-n-m}$ are model-dependent parameters which are to be constrained by experiment and observation. Typically, only one of these parameters is considered to be nonvanishing to single out a model, for instance $a_{0,1,2}$ is associated to the bicrossproduct basis of the $\kappa$-Poincar\'e algebra \cite{Kowalski-Glikman:2002iba}, while $a_{0,0,3}$ relates to LIV-models \cite{Kostelecky:2008ts} (and a different representation of the $\kappa$-Poincar\'e algebra \cite{Girelli:2006fw,Amelino-Camelia:2014rga}). At this level, however, we intend to be as general as possible.

Being dependent on energies and momenta, Eq. \eqref{mdr1} can only be valid in a particular reference frame. In order to give meaning to a dispersion relation of this kind, then, it is necessary to provide means of defining these quantities. This is all the more important in the curved-spacetime setting with underlying metric $g_{\mu\nu}(x)$ which we aspire to entertain in the present paper. To generalise Eq. \eqref{mdr1} to this context, we have to introduce a deformation vector $d$ which is dimensionless, timelike and normalised, thus satisfying the equation $d^{\mu}d^\nu g_{\mu\nu}=-1.$ It is through this very procedure that an evident tension arises with respect to the foundational principle of local Lorentz invariance. Nevertheless we continue to speak of $\mathcal{C}$ a Casimir operator to bear in mind that the underlying momentum space may be maximally symmetric \cite{Kowalski-Glikman:2002oyi}, thus allowing for 10 isometries such that Lorentz invariance is deformed rather than being explicitly violated. 

Given the vector $d,$ we can foliate spacetime into the part parallel to it and the hypersurfaces making out its complement. The projector on those spacelike hypersurfaces reads
\begin{equation}
    g_{\perp\nu}^{~\mu}\equiv\delta^\mu_\nu+d^\mu d_\nu
\end{equation}
such that $g_{\perp\nu}^{~\mu} d^\nu=g_{\perp\mu}^{~\nu} d_\nu=0.$ In this vein, we measure energy along and momentum normally to $d$ such that
\begin{align} k_{\perp\mu}=g_{\perp\mu}^{~\nu} k_\nu,&& E=k_dc\equiv d^\mu k_\mu c=\sqrt{M^2c^2+k_{\perp}^2}c+\mathcal{O}(\ell),
\end{align}
where norm of the spatial momentum can be expressed in terms of the induced metric $g_\perp^{\mu\nu}$ as
\begin{equation}
    k_\perp=\sqrt{g^{\mu\nu}_\perp k_\mu k_\nu}.
\end{equation}
In terms of $k_\perp,$ we can generalise Eq. \eqref{mdr1} to arbitrary spacetime frames, thus obtaining
\begin{equation}
    M^2c^2=g^{\mu\nu}k_\mu k_\nu+\ell\sum_{n=0}^2\sum_{m=0}^{3-n} a_{n,m,3-n-m}(Mc)^n (M^2c^2+k_\perp^2)^\frac{m}{2} k_\perp^{3-n-m}.\label{eqn:mdr1CST}
\end{equation}

Dispersion relations of this type introduce alterations to particle trajectories that scale proportionally with the product of mass/energy/momentum and the Planck length. Conventional wisdom, then, has it that the main amplifier required to meaningfully constrain the model parameters $a_{n,m,3-m-n}$ consists in high energies. For example, this phenomenon gives rise to a distinct ``in-vacuo dispersion'' of particles, observable through time delays in the arrival of massless particles emitted from cosmological sources \cite{HESS:2011aa,MAGIC:2017vah,MAGIC:2020egb}, as well as the occurrence of threshold effects \cite{HAWC:2019gui}. The in-vacuo dispersion can be interpreted as resulting from the interaction between a particle and the quantum spacetime degrees of freedom. Several fundamental theories of quantum gravity, such as loop quantum gravity \cite{Gambini:1998it,Amelino-Camelia:2016gfx} and non-critical string theory \cite{Amelino-Camelia:2016gfx}, provide circumstantial evidence that supports this notion. 

In addition to high-energy phenomena acting as prominent amplifiers of Planck-scale effects, the IR regime holds notable significance for experiments conducted with exceptional precision (where the precision can lead to Planck scale sensitivity even in the absence of an exceptionally large amplifier), such as table-top experiments \cite{Carney:2018ofe}. Cold atoms serve as an intriguing example wherein Lorentz violation of this nature can be constrained \cite{Amelino-Camelia:2009wvc}. Traditionally, findings within this regime have been derived based on simplistic expectations regarding the modification of quantum observables. However, the present study takes a more deliberate approach by carefully investigating the ramifications of assuming an MDR on IR quantum mechanics. Yet, considering quantum dynamics on curved spacetime, the very definition of the nonrelativistic regime harbours a number of intricacies, which shall be thoroughly addressed in the subsequent section.


\section{The subtleties of the nonrelativistic limit}

There is a long history of attempts at obtaining nonrelativistic limits of underlying relativistic theories and possible post-Newtonian corrections in classical (for a recent introduction see section 3 of \cite{Will:2014kxa}) as well as quantum mechanics (see \eg 
\cite{Lammerzahl:1995zz,Nicolini:2012eu,Schwartz:2018pnh}). This is to say that, while appearing straight-forward, an expansion in $1/c$ can be subtle. Most of these subtleties revolve around $c$ being dimensionful. In other words, before expanding in $1/c,$ we need to make clear which quantities are to be compared to the speed of light. In the following, we summarise some of the subtleties and ambiguities that appear along the way and provide the perspective on the\textit{}m used in the present work.

\subsection{Breaking covariance}\label{sec:cov}

Intuitively, in the nonrelativistic regime, three-velocities of the described objects, in some sense, need to be small when compared to the speed of light. The same goes for spatial momenta with respect to the mass in the Hamiltonian formalism. However, by Einstein's principle of relativity there is no absolute notion of velocity or momentum. Thus, when we assert that some velocity is small, this is a coordinate-dependent statement. Therefore, from the moment on that we truncate a series in $1/c$ at finite order, we break covariance. However, when considering nonrelativistic physics, this is a feature rather than a bug. In Newtonian mechanics, time is absolute, while in relativity it is not. Thus, we better provide a preferred frame which can tell us which exactly is the Newtonian absolute time and who (or what) is moving slowly.

In other words, before being able to actually perform the calculation, it is necessary to single out a timelike curve $\gamma$, relative to which we want to expand our theory. Along this curve, we can define a tangent vector $\dot{\gamma}^\mu=\D\gamma^\mu/\D\lambda,$ with a curve parameter with dimensions of length $\lambda,$\footnote{Without loss of generality, we assume that the general coordinate system which marks our starting point, indicated by the Greek letter $\mu$, consists of quantities of units of length. Then, the metric $g_{\mu\nu}$ is dimensionless. This will cease to be the case in the adapted set of coordinates introduced below where the time-coordinate $t$ will indeed have units of time.} and where we impose the normalisation $\gamma^\mu\gamma^\nu g_{\mu\nu}=-1.$ Given this vector, we may foliate the background spacetime as
\begin{equation}
    g_{\mu\nu}=h_{\mu\nu}-\dot\gamma_\mu\dot\gamma_\nu,\label{partition}
\end{equation}
where $h_{\mu\nu}$ describes the induced metric on spacelike hypersurfaces to which $\dot{\gamma}^\mu$ is normal, \ie $\dot{\gamma}^\mu h_{\mu\nu}=0$. To wit, the form $\dot{\gamma}_\mu\D x^\mu$ defines the direction of time which is supposed to become global in the expansion. 

While the four-dimensional notation is appropriate at the relativistic level, when considering nonrelativistic physics, it is helpful to revert to adapted coordinates $(ct,x^i)$ which are defined by
\begin{equation}
    \dot\gamma_\mu\D x^\mu\equiv-N\D t,\label{eqn:defAdCoord}    
\end{equation} 
with the normalisation factor $N,$ namely the lapse function. Then, the adapted-coordinate time $t$ can also be understood as a curve parameter along $\gamma$, defined up to time-reparameterisation invariance (which we will take advantage of in Secs. \ref{sec:CanQant} and \ref{sec:ModKf}). Given a general background geometry, it can be shown that the line element can be expressed in terms of ADM-variables \cite{Arnowitt:1959ah,Arnowitt:1960es} (see chapter 12 of \cite{Padmanabhan:2010zzb} for a recent introduction), \ie $N$ as well as the shift vector $N^i$ and the induced metric $h_{ij},$ as
\begin{equation}
    \D s^2=-N^2c^2\D t^2+h_{ij}(N^ic\D t+\D x^i)(N^jc\D t+\D x^j).\label{ADMmetric}
\end{equation}
Here the lapse function and the shift vector are pure gauge degrees of freedom. They are fixed by the choice of curve $\gamma.$ It is in these adapted coordinates that the overall dependence on $1/c$ is manifest, thus significantly simplifying the expansion.

We are now in the position to understand the meaning of the term ``small velocities''. In adapted coordinates, we may use the soon-to-be global time as curve parameter such that
\begin{equation}
    U^\mu\equiv\frac{\D x^\mu}{\D (ct)}=\left(1,c^{-1}\frac{\D x^i}{\D t}\right)\equiv \left(1,\frac{\dot{x}^i}{c}\right).\label{eqn:FourVel}
\end{equation}
It is the quantity $\dot{x}^i/c$ that we have to expand in. Similarly, on the level of momenta $p_i/M c\ll 1,$ with the mass of the considered object $M.$ 

A closer look at Eq. \eqref{eqn:FourVel} tells us that an analogous expansion has to affect vector and tensor fields. Throughout this paper, we will indicate this series as
\begin{align}
        \mathcal{T}=\sum_{n=0}^\infty c^{-n}\mathcal{T}^{(n)},\label{BGFieldExp}
\end{align}
where the symbol $\mathcal{T}$ stands for any general tensor field content.\footnote{For example, this includes the metric either in its covariant formulation $g_{\mu\nu}$ or the canonical one in terms of $N,$ $N^i$ and $h_{ij},$ but also additional (quantum-gravity inspired or electromagnetic) fields.}

The analogue of Eq. \eqref{eqn:FourVel} for general vectors $V^\mu$ has particular consequences at lowest order, \ie in the limit $c\rightarrow\infty.$ In adapted coordinates such a vector may be written as
\begin{equation}
        V_{(0)}^\mu\equiv\lim_{c\rightarrow \infty}\frac{\D\gamma_V^\mu}{\D t}=\lim_{c\rightarrow \infty}\left(1,c^{-1}\frac{\D \gamma_V^i}{\D t}\right)= \left(1,0\right),\label{eqn:LimVec}
\end{equation}
where $\gamma_V$ denotes the curve to which $V^\mu$ is tangent, and we used the reparametrisation invariance of the curve to set $\D\gamma_V^0/\D t=1.$ As a result, in this limit every vector is either timelike, and possesses vanishing spacelike components or it vanishes altogether. Thus, in adapted coordinates, every spatial vector such as, for example, $N^i$ is vanishing at lowest order, \ie $N^i_{(0)}=0$ such that $\dot\gamma^\mu_{(0)}=N^{-1}_{(0)}\delta^\mu_0.$ On the level of scalar quantities, we may formulate this fact as
    \begin{equation}
        \lim_{c\rightarrow\infty}V^\mu V^\nu g_{\mu\nu}=\lim_{c\rightarrow\infty}\left[g_{00}+c^{-1}\frac{\D \gamma_V^i}{\D t}g_{0i}+c^{-2}\frac{\D \gamma_V^i}{\D t}\frac{\D\gamma_V^j}{\D t}g_{ij}\right]= -N_{(0)}^2,\label{ZeroOrderTimelike}
\end{equation}
which is negative, implying that the vector is either timelike or equal to zero. Geometrically, this can be understood as the light-cones making out the entire manifold as $1/c$ approaches $0$ -- spacelike-separated events are an infinite distance apart. 

For normalised timelike vectors $v^\mu$, we may go even further. In this case, we obtain
    \begin{equation}
        v^\mu v^\nu g_{\mu\nu}\simeq -\left(v_{(0)}^0\right)^2N_{(0)}^2-2c^{-1}\left(v^0_{(1)}v^0_{(0)}N_{(0)}^2+(v^0_{(0)})^2N_{(0)}N_{(1)}\right)=-1,\label{ZeroCompNormed}
\end{equation}
which can only be solved by $v^\mu_{(0)}=\pm\dot\gamma^\mu_{(0)}$ and $v_{(1)}^0=\mp N_{(1)}/N_{(0)}^2.$ Thus, to lowest order, any normalised timelike vector is either aligned or antialigned with the global time and the absolute value of the time-component of any timelike unit vector is equal up to second-order corrections. In particular, this observation applies to the Lorentz-violating vector $d^\mu$ introduced in the preceding section.

In a nutshell, there is a lot of information that can be derived from the necessity of breaking covariance alone. This, however, is not the only subtlety of the nonrelativistic limit in curved geometries. 

\subsection{External scales}\label{sec:ExtSca}

As we have seen in the preceding subsection, all background tensor fields are to be expanded in $1/c.$ This implies that, in general, we are not just comparing velocities to the speed of light. External fields may contain additional scales. For example, the speed of light prominently features in the coupling of Einstein's field equations
\begin{equation}
        G_{\mu\nu}=\frac{8\pi G}{c^4}T_{\mu\nu},\label{eqn:EFE}
\end{equation}
with the Einstein tensor $G_{\mu\nu},$ the gravitational constant  $G$ and the stress-energy tensor $T_{\mu\nu}$ (for a thorough introduction consult \cite{Carroll:2001ws}). Thus, blindly expanding in $1/c$ implicitly requires gravity to be weak. However, depending on the process that is to be described, this may not be the most sensible way to proceed -- the nonrelativistic limit does not necessarily preclude strong gravity. This kind of ambiguity is to be expected when expanding in dimensionful parameters like $1/c.$ 

As the present paper is directed towards low-energy phenomenology, we choose gravitational corrections to be negligible at lowest order, \ie generically presume that $N_{(0)}=1.$

\subsection{Quantum-gravity inspired corrections}\label{sec:QGCorr}

External scales cannot only arise when coupling to classical gravity or other forces. The quantum-gravity inspired corrections to relativistic particle dynamics we introduce in sections \ref{sec:CanQuantMDR} and \ref{sec:ModKfMDR} themselves depend on the Planck length $\ell$. The Planck length, in turn, is defined as
\begin{equation}
    \ell\equiv\sqrt{\frac{\hbar G}{c^3}}\sim 10^{-35}m,
\end{equation}
which clearly depends on $c.$ Similarly, in the $1/c$-expansion of the correction term, there will be terms behaving like positive powers of $c.$ In the strict limit of $c\to\infty$ these would become infinitely large.\footnote{Similar issues have been encountered when taking the classical limit of minimal-length quantum mechanics \cite{Casadio:2020rsj}. They have been elaborated upon and discussed further in \cite{Bosso:2023aht}.} In the phenomenological context, however, this is irrelevant because the Planck length itself is exceedingly small and, independently of the appearance of the speed of light, serves as perturbative parameter when compared to other length (or inverse-energy) scales. Bear in mind that it is not required to apply the strict limit of $c\to\infty.$ In reality, in the nonrelativistic regime, the speed of light is not infinite, it is just very large. In particular it can be compensated for by other scales. 

Henceforth, we refrain from expanding the phenomenological corrections to the same degree in $1/c$ as we do with other terms. Instead, we solely focus on their leading contribution in $1/c$, coupled with the Planck length $\ell$, an independent expansion in powers of which is understood. By adopting this approach, we recognise the importance of singling out the leading effects that manifest within the interplay of these distinct factors.

\subsection{Time-dependent backgrounds in quantum mechanics}\label{sec:TimeDepBack}

Below, we intend to obtain nonrelativistic quantum dynamics in two different ways -- by virtue of canonical quantisation of classical dynamics and as a limit of the covariant Klein-Gordon equation (and its generalisations). 
When confronted with a background geometry that lacks a timelike Killing vector field, both of these approaches encounter their respective limitations. 

In short, the issue lies in the fact that the natural Hilbert-space scalar product on surfaces of constant time (\ie normal to $\D t$), be it derived from the Klein-Gordon scalar product \cite{Mostafazadeh:2006ug}
\begin{equation}
        \langle{\Psi|\Phi\rangle}_\text{KG}=\int\D\Sigma^\mu_{t_0} \left[\Psi^*\nabla_\mu\Phi-(\nabla_\mu\Psi)^*\Phi\right],\label{eqn:KGMeas}
\end{equation}
or created by canonical quantization according to geometric calculus \cite{Pavsic:2001ih}, contains the volume form on these hypersurfaces $\Sigma_t$ \cite{DeWitt:1952js,Wagner:2021bqz,Wagner:2021luc}, \ie
\begin{equation}
    \langle\psi|\phi\rangle=\int_{\Sigma_{t}}\D^d x\sqrt{h}\psi^*\phi,\label{eqn:GenMeas}
\end{equation}
where $h=\det h_{ij}$ in adapted coordinates. If the volume measure $\sqrt{h}$ appearing in this integral changes with time, the unitarity of the resulting quantum theory ceases to be obvious \cite{Kiefer:1990pt} just because Hilbert-space elements at infinitesimally different times cannot be identified anymore \cite{Schwartz:2018pnh}. In a similar vein, the very notion of single-particle quantum theory becomes slightly problematic in evolving backgrounds due to particle creation. 

In the main body of the paper, we will assume that the spatial part of the metric is constant, \ie in adapted coordinates $\partial_0h_{ij}=0.$

\section{Canonical quantisation of nonrelativistic dynamics}\label{sec:CanQant}

This section focuses on the first of two methods employed to derive the deformed Schrödinger equation from an MDR. We begin by considering the nonrelativistic limit at the classical level. Next, we quantise the resulting classical theory using canonical techniques, following the established approach to quantisation on curved spaces originally proposed by deWitt \cite{DeWitt:1952js}.

In general, starting at a dispersion relation of the kind given in Eq. \eqref{GenDisRel}, we have to choose a slicing along which we define the Hamiltonian which governs single-particle dynamics. Conveniently, for the nonrelativistic limit, we employ the adapted coordinate system $x^\mu=(ct,x^i)$ introduced in Sec. \ref{sec:cov}. Provided the ADM-decomposition of the metric \eqref{ADMmetric}, Eq. \eqref{GenDisRel} can be solved for $k_0.$ Then, the dynamics of a particle along the worldline $\gamma$ is governed by the Hamiltonian
\begin{equation}
    H=c k_0 (k_i,M,\Phi),
\end{equation}
where $\Phi$ stands for general external field content (like the metric or an electromagnetic field). In other words, the Hamiltonian provides the energy as measured along $\gamma$. This function is to be expanded in $1/c,$ first for Riemannian backgrounds, and subsequently for a deformations thereof corresponding to the MDR in Eq. \eqref{eqn:mdr1CST}.

\subsection{Riemannian spacetime}

In order to describe motion in curved spacetime, the special relativistic dispersion relation has to be enhanced. As a result, the dynamics of a relativistic particle on a curved spacetime of four-momentum $k_\mu$ is to satisfy the constraint
\begin{equation}
    \mathcal{C}_0=-g^{\mu\nu}(x)k_\mu k_\nu=M^2c^2.\label{GRCasimir}
\end{equation}
Demanding that the energy of the particle be positive and using Eq. \eqref{ADMmetric}, we find the Hamiltonian
\begin{equation}
    H_0=NMc^2\sqrt{1+\frac{h^{ij}k_ik_j}{(Mc)^2}}+N^ik_ic.\label{GRHam}
\end{equation}
We expand this function in velocities, thus assuming that $h^{ij}k_ik_j\ll M^2c^2,$  as well as the background fields as required by Eq. \eqref{BGFieldExp}.

The nonrelativistic limit in and of itself consists in letting the speed of light tend to infinity. This is only a well-defined procedure if the terms in the Hamiltonian proportional to positive powers of $c$ do not contribute to the equations of motion, \ie amount to constants. In order for this condition to be satisfied, the quantity $N_{(1)}$ has to be constant such that it can be set equal to zero by a constant reparameterisation of the time-coordinate.

As a result, the unperturbed three-dimensional tensor $h_{ij}^{(0)}$ becomes the effective metric on the resulting hypersurfaces. In the following, the induced metric and its inverse $h^{ij}_{(0)}$ are used to raise and lower spatial indices. Neglecting the trivial constant part, the nonrelativistic Hamiltonian amounts to
\begin{equation}
    H_{0,\text{NR}}=\frac{h^{ij}_{(0)}}{2M}\left(k_i-MA^{\rm g}_i\right)\left(k_j-MA_j^{\rm g}\right)+\phi_{\rm g}M,\label{eqn:HNRBefCan}
\end{equation}
with the gravitomagnetic vector potential $A^{\rm g}_i=-N_i^{(1)}$ and the gravitoelectrostatic potential $\phi_{\rm g}=(N_{(2)}-N^i_{(1)}N_j^{(1)}/2)$ (for further details, refer to \cite{Petruzziello:2021vyf}). The resemblance of this Hamiltonian to the one describing a charged particle in the presence of an electromagnetic field is evident. Here, the r\^ole of the charge is assumed by $M$. 

Following deWitt's approach, the quantisation of the dynamics of single particle on a curved background is preferably obtained by choosing the Hilbert-space measure to be provided by the invariant volume-form $h\D^d x,$ where $h=\det h_{ij}$ \cite{DeWitt:1952js,Wagner:2021bqz,Wagner:2021luc}. Correspondingly, the wave functions $\psi\in\mathcal{H}$ are scalars and so is the representation of the Hamiltonian. Following the principles of geometric calculus \cite{Pavsic:2001ih}, this representation is most straight-forwardly provided in terms of covariant derivatives $\nabla_i$ compatible with the induced metric. Including the minimal coupling of an external magnetic field obtained by enforcing local $U(1)$ invariance of the wave function (\ie $\hat{k}_i\to \hat{k}_i -eA_i,$ $\hat{H}\to \hat{H}+e\phi$), we obtain the operator-equivalent of the Hamiltonian
\begin{equation}
    \hat{H}_{0,\text{NR,EM}}\psi=\left[-\frac{h^{ij}}{2M}\left(\nabla_i-ieA_i-iMA^{\rm g}_i\right)\left(\nabla_j-ieA_j-iMA^{\rm g}_j\right)+e\phi_e+M\phi_{\rm g}\right]\psi.\label{HNREMRep}
\end{equation}
Thus, in the nonrelativistic limit we observe a particle moving on a curved $d$-dimensional geometry which is charged under a $U(1)_e\times U(1)_M$-symmetry with charges $e$ and $M.$ 

This finding is in line with \cite{Schwartz:2018pnh}, nevertheless corresponding to a generalisation thereof because in the present paper the spatial metric is not assumed to be flat. Equation \eqref{HNREMRep} is to be extended to MDRs in the subsequent subsection.

\subsection{Modified dispersion relations}\label{sec:CanQuantMDR}

In contrast to the Riemannian case, solving Eq. \eqref{GenDisRel} for $k_0$ is highly nontrivial. However, we are only interested in solutions to order $\ell.$ Therefore, we can approximately reduce Eq. \eqref{GenDisRel} such that
\begin{equation}
    \mathcal{C}_0=M^2c^2-\left.\ell\sum_{n=0}^2\sum_{m=0}^{3-n} a_{n,m,3-n-m}(Mc)^n k_\perp^m k_d^{3-n-m}\right|_{k_0=H_0/c}= (Mc)^2-\delta\bar{\mathcal{C}},\label{DefDeltaC}
\end{equation}
where we abbreviated $\delta\bar{\mathcal{C}}\equiv\delta \mathcal{C}|_{k_0=H_0/c}.$ Consequently, we obtain the deformed Hamiltonian
\begin{equation}
    H_\ell=NMc^2\sqrt{1+\frac{h^{ij}k_ik_j-\delta\bar{\mathcal{C}}}{(Mc)^2}}+N^ik_ic\simeq H_0-\frac{N\delta\bar{\mathcal{C}}}{2M}\left(1+\frac{h^{ij}k_ik_j}{(Mc)^2}\right)^{-1/2}.\label{DefRelHamClass}
\end{equation}
As argued in Sec. \ref{sec:QGCorr}, only the leading-order correction in $1/c$ is of interest to us. Bear in mind, however, that relevant contributions is not allowed to amount to a total derivative so as to contribute nontrivially to the dynamics. In particular, it has to be dependent on positions and/or momenta.

In order to evaluate the leading contribution of $\delta \bar{C},$ we have to study the projections of the four-momenta respective to the slicing carved out by $d^\mu,$ \ie $k_d$ and $k_\perp$ under the replacement $k_0\rightarrow H_0/c.$ 
This is most effectively done by expressing the Lorentz-violating vector in terms of its deviation from the global time such that
\begin{equation}
    d^\mu=\dot\gamma^\mu-\mathcal{A}^\mu,\label{eqn:DevLorVio}
\end{equation}
where we may normalise in accordance with Eq. \eqref{ZeroCompNormed} such that
\begin{align}
    \mathcal{A}^\mu_{(0)}=\mathcal{A}^0_{(1)}=0,&&\mathcal{A}^0_{(2)}=\frac{\mathcal{A}_{(1)}^i\mathcal{A}_{(1)}^jh_{ij}^{(0)}}{2N_{(0)}}.\label{eqn:DefNormCond}
\end{align}
Applying these conditions and expanding to second order, we obtain the momentum projected to the hypersurfaces
\begin{equation}
k_\perp|_{k_0=H_0/c}\simeq\sqrt{h^{ij}_{(0)}\left(k_i-M\mathcal{A}^{(1)}_i\right)\left(k_j-M\mathcal{A}^{(1)}_j\right)}.\label{EffPperp}
\end{equation}
This quantity deviates from the magnitude of the momentum inasmuch as it contains the quantity $\mathcal{A}_i^{(1)}.$ Having units of velocity, this vector quantifies the relative velocity between the reference system for the nonrelativistic limit and the deformation vector.

Defining $k^\mathcal{A}_i\equiv k_i-M\mathcal{A}_i$ and $k_\mathcal{A}^2\equiv h^{ij}_{(0)}k^\mathcal{A}_ik^\mathcal{A}_j,$ we can finally express the corrected nonrelativistic Hamiltonian as
\begin{align}
    H_{\ell,\text{NR}}\simeq H_{\text{NR}}-\frac{\ell}{2M}\sum_{n=0}^2\sum_{m=0}^{3-n}a_{n,m,3-n-m}(Mc)^{3-m}k_\mathcal{A}^{m}\left[1+\frac{3-n-m}{2Mc}k_\mathcal{A}^2\right]\label{HlNRp}.
\end{align}
While it appears that we display not only the leading but also the next-to-leading order in $1/c$ here, this is rooted in the fact that the leading contribution in models with $m=0$ is constant. A constant addition to a Hamiltonian, however, has no influence on the equations of motion.

To make this manifest, we parameterise the deformed nonrelativistic Hamiltonian differently such that it becomes
\begin{equation}
    H_{\ell,\text{NR}}\simeq H_{\text{NR}}-\frac{\ell}{2M}\sum_{n=1}^3\xi_n(Mc)^{3-n}k_\mathcal{A}^{n}.\label{HlNR}
\end{equation}
The newly introduced dimensionless parameters $\xi_n$ are clearly more suited to the nonrelativistic dynamics. They are related to the original parameterisation as
\begin{equation}
    \xi_1=a_{1,1,1}+a_{2,1,0}+a_{0,1,2},\quad\xi_2=a_{1,2,0}+a_{0,2,1}+\frac{1}{2}a_{2,0,1}+a_{1,0,2}+\frac{3}{2}a_{0,0,3},\quad\xi_3=a_{0,3,0}.\label{eqn:xis}
\end{equation}
We are left with the task of canonically quantising the corrections. By analogy with Sec. \ref{sec:CanQant}, this can be done by representing momenta as gauge-covariant derivatives when acting on position eigenstates. The Hamiltonian thus acts on wave functions $\psi$ as
\begin{align}
    \hat{H}_{\ell,\text{NR},\text{EM}}\psi=\hat{H}_{0,\text{NR},\text{EM}}\psi-\frac{\ell}{2M}\sum_{n=1}^3\xi_n(Mc)^{3-n} \left[-h^{ij}_{(0)}(\nabla_i-iM\mathcal{A}_i-ieA_i)(\nabla_j-iM\mathcal{A}_j-ieA_i)\right]^{n/2}\psi.\label{eqn:DefCanSchroEl}
\end{align}
As the corrections are functions of one operator only, namely the Laplacian with an insertion of the field $\mathcal{A}_i,$ they are automatically symmetric with respect to the ordinary volume measure $h\D^d x.$ Notice that the gravitomagnetic potential $A_i^{\rm g}$ only contributes to the unperturbed Hamiltonian, and does not appear in the correction term. In the subsequent section, this result is complemented with the deformed Schr\"odinger equation derived from relativistic field theory.

\section{Nonrelativistic limit of modified Klein-Gordon equation}\label{sec:ModKf}

An alternative avenue to derive nonrelativistic quantum dynamics from its relativistic counterpart is by applying the nonrelativistic limit to the single-particle sector of quantum field theory. This approach relies on a somewhat heuristic methodology, as it necessitates the division of modes into positive and negative frequencies, a task that generally eludes perturbative treatment within the $1/c$ expansion, refer to \cite{Schwartz:2018pnh} for more details. As discussed in Sec. \ref{sec:TimeDepBack}, due to particle creation this is certainly an issue in time-dependent backgrounds.

In order to describe the dynamics of a single spinless particle, charged with respect to $U(1)$ gauge field, we consider the covariant Klein-Gordon equation as well as generalisations thereof, \ie in adapted coordinates
\begin{equation}
    \left[\hat{\mathcal{C}}(-iD_0,-iD_i,M)-M^2\right]\varphi=0\label{ModKG}.
\end{equation}
Here, $\varphi$ is the scalar field, and $D_\mu$ stands for the gauge-covariant derivative including the $U(1)$-gauge field $A_\mu$ and the charge $e,$ \ie
\begin{equation}
    D_\mu=\nabla_\mu-ieA_\mu.
\end{equation}
The covariant derivative $\nabla_\mu$ is based on the Levi-Civita connection $\Gamma^\rho_{\mu\nu}$ compatible with the background metric $g_{\mu\nu}$. 

Inspired by an approach developed in \cite{Kiefer:1990pt,Lammerzahl:1995zz,Giulini:2012zu,Schwartz:2018pnh}, we apply a WKB-like expansion to Eq. \eqref{ModKG}. In this vein, we assume the scalar field to be given as
\begin{equation}
    \varphi=e^{-iMc\lambda}\psi=e^{-iMc\lambda}\sum_{n=0}^{\infty}\frac{\psi_n}{c^n},\label{FieldExp}
\end{equation}
with some scalar function with units of length $\lambda,$ assumed to be uncharged under $U(1),$ while the field $\psi=\sum_{n=0}^{\infty}c^{-n}\psi_n$ carries the charge $e.$ 
From the field equation, it is possible to derive the effective equation of motion for the Schr\"odinger field $\psi_0$. First, this procedure will be followed for general Riemannian geometries, to be subsequently generalised to deformed Casimir operators.

\subsection{Riemannian spacetime}

On a curved spacetime characterised by the metric $g^{\mu\nu}(x)$ and considering a $U(1)$-gauge field, the mass Casimir equals the covariant d'Alembertian 
\begin{equation}
    \hat{\mathcal{C}}_0(-i\nabla_\mu,A_\mu)=g^{\mu\nu}D_\mu D_\nu.
\end{equation}
Plugging in the ansatz in Eq. \eqref{FieldExp}, the corresponding Klein-Gordon equation reads
\begin{equation}
    \left[-M^2c^2(g^{\mu\nu}\nabla_\mu \lambda\nabla_\nu \lambda+1)-iMc\left(\Box \lambda+2g^{\mu\nu}\nabla_\mu \lambda D_\nu\right)+g^{\mu\nu}D_\mu D_\nu\right]\psi=0,\label{KG}
\end{equation}
with the d'Alembertian $\Box=g^{\mu\nu}\nabla_\mu \nabla_\nu.$ This equation is to be expanded in powers of $1/c.$ In accordance with Eq. \eqref{BGFieldExp}, the involved background fields, \ie the metric $g_{\mu\nu}$ and the gauge field $A_\mu,$ necessarily have to be expanded as well.

At order $c^2,$ the covariant Klein-Gordon equation yields the condition
\begin{equation}
    -g^{\mu\nu}_{(0)}\partial_\mu\lambda\partial_\nu\lambda=1.
\end{equation}
This equation implies two points. 

\begin{itemize}
    \item First, 
    it characterises $\lambda$ as the proper distance (or rescaled proper time) covered along a timelike curve on the geometry compatible with the metric $g_{\mu\nu}^{(0)}$, \ie $\lambda=c\tau_{(0)}.$ Thus, the solution \eqref{FieldExp} indeed corresponds to a single particle moving along this curve.
    \item Second, the timelike form $\partial_\mu\lambda$ is normalised at zeroth order. Thus, by Eq. \eqref{ZeroCompNormed}, it is aligned with the global time, \ie $\dot{\gamma}^{(0)}\D x^\mu=\D\lambda.$ In adapted coordinates, we may express this condition as $\D\lambda=-N_{(0)}c\D t,$ which suffices to foliate the background spacetime into evolving hypersurfaces of constant time as in Eq. \eqref{partition}. 
\end{itemize}

When further expanding Eq. \eqref{KG} in powers of $1/c,$ it is important to bear in mind that $g^{\mu\nu}_{(0)}\dot\gamma_\nu^{(0)}\nabla_\mu^{(0)}$ in adapted coordinates is to become a derivative with respect to the time coordinate, thus being accompanied by a factor of $1/c.$ This will cease to be the case at higher order, where the vector $g^{\mu\nu}\dot{\gamma}^{(0)}_\mu$ gains spacelike components. Similarly, the timelike component of the electromagnetic potential $\dot{\gamma}^{(0)}_{\mu}A^\mu,$ in general, goes at most as $1/c$. At order $c$ Eq. \eqref{KG} then reduces to a complex constraint on the background metric. Thus, we obtain two consistency conditions for the existence of solutions of the form \eqref{FieldExp}, \ie
\begin{equation}
    g^{\mu\nu}_{(1)}\dot{\gamma}_\mu^{(0)}\dot{\gamma}_\nu^{(0)}=K_{(0)}=0,\label{eqn:condextcur}
\end{equation}
with the trace of the extrinsic curvature of the spacelike hypersurfaces $K\equiv-h^{\mu\nu}\nabla_\mu\dot{\gamma}_\nu.$ Consequently, similarly to the classical case $N_{(1)}=0$  (\cf the reasoning below Eq. \eqref{GRHam}), and the extrinsic curvature vanishes at lowest order. This is the logical consequence of there only being timelike vectors at that level.  In adapted coordinates, the extrinsic curvature reads \cite{Padmanabhan:2010zzb}
\begin{equation}
    K=\frac{h^{ij}}{2N}\left(2\nabla_iN_j-c^{-1}\partial_0h_{ij}\right).\label{extrinsicCurvature}
\end{equation}
Hence, $K_{(0)}=0$ is trivially satisfied. Finally, at order $(1/c)^0$, we obtain the effective Schrödinger equation
\begin{equation}
    i\partial_0\psi=\left[-\frac{h^{ij}}{2M}\left(\nabla_i-ieA_i-iM A^{\rm g}_i\right)\left(\nabla_j-ieA_j-iM A^{\rm g}_j\right)+M\phi_{\rm g}+e\phi_e\right]\psi,\label{eqn:DensitySchrodinger}
\end{equation}
where the gravitational and magnetic potentials are defined as below Eq. \eqref{eqn:HNRBefCan}, and where we removed the sub- and superscripts indicating the orders to avoid unnecessary cluttering. 

As $N_{(0)}=1$ and $N^i_{(0)}=1$, the spatial part of the covariant derivatives $\nabla_i^{(0)}$ is compatible with the induced metric $h^{ij}_{(0)}.$ Therefore, the vector operator $-i\nabla_i$ is Hermitian with respect to the effective Klein-Gordon measure given in Eq. \eqref{eqn:KGMeas}.  As in Sec. \ref{sec:CanQuantMDR}, Eq. \eqref{eqn:DensitySchrodinger} will now be generalised to MDRs.

\subsection{Modified dispersion relations}\label{sec:ModKfMDR}

Modifications of the mass Casimir as in Eq. \eqref{eqn:mdr0} imply deformations to the Klein-Gordon equations in accordance with Eq. \eqref{ModKG}. In this sense, we may interpret the four-momentum as a gauge-covariant derivative. Define the operator-valued counterparts of $k_\perp$ and $k_d$ as
\begin{align}
    \hat{k}_\perp^2\varphi=&-D_\mu g^{\mu\nu}_\perp D_\nu\varphi,\label{eqn:pperpop}\\
    \hat{k}_d\varphi=&\frac{i}{2}\{d^\mu, D_\mu\}\varphi =\sqrt{M^2c^2+\hat{k}_\perp^2}\psi+\mathcal{O}(\ell),\label{eqn:ptop}
\end{align}
the anticommutator $\{,\}$ was introduced to render the operator $\hat{k}_d$ symmetric with respect to the Klein-Gordon measure (provided in \eqref{eqn:KGMeas}). Consequently, $[\hat{k}_d,\hat{k}_\perp]=0$ at order $\ell^0.$ This implies that the pseudo-Riemannian Casimir is modified as (\cf \cite{Kowalski-Glikman:2003qjp,Kruglov:2012sf,Majhi:2013koa,Franchino-Vinas:2022fkh})
\begin{equation}
    \hat{\mathcal{C}}_\ell=\hat{\mathcal{C}}_0 +\frac{\ell}{2}\sum_{n=0}^2\sum_{m=0}^{3-n} a_{n,m,3-n-m}(Mc)^n\hat{k}_\perp^{m}\hat{k}_d^{3-n-m}=\hat{C}_0+\delta\hat{C}.
\end{equation}
The model coefficients $a_{n,m,3-m-n}$ are in one-to-one correspondence to the ones in Eq. \eqref{mdr1}.

In the following, we aspire to obtain the leading corrections to the action of $\delta\hat{C}$ on the scalar. In this vein, it is crucial to recognize that the phase factor satisfies $D_i^{(0)}\lambda=0$ and $D_0^{(0)}\lambda=-c$. Thus, to lowest order the phase is an eigenfunction of the covariant time-derivative operator
\begin{equation}
    iD_0^{(0)}e^{iMc\lambda}=Mc^2e^{iMc\lambda}.\label{eqn:EigFunc}
\end{equation}
As a result, the combination $D_0^{(0)}\varphi$ is of order $c^2$, while $D_i^{(0)}\varphi$ is of order one. Furthermore, at order $\ell$ the time-derivative of the wave function can be approximated by applying Eq. \eqref{eqn:DensitySchrodinger} to it. Then, using Eq. \eqref{eqn:DefNormCond} and parameterising the deviation of the Lorentz-violating vector from the tangent vector of the curve $\gamma$ in accordance with Eq. \eqref{eqn:DevLorVio}, we obtain the leading-order representations
\begin{align}
    e^{-Mc\lambda}\hat{k}_\perp^2\varphi\simeq&-h^{ij}_{(0)}\left(D^{(0)}_i-iM\mathcal{A}_i^{(1)}\right)\left(D^{(0)}_j-iM\mathcal{A}_j^{(1)}\right)\psi\equiv \hat{k}_\mathcal{A}^2\psi,\\
    e^{-Mc\lambda}\hat{k}_d\varphi\simeq&\left(Mc+\frac{\hat{k}_\mathcal{A}^2}{2Mc}\right)\psi.
\end{align}
With these expressions in mind, we can obtain the corrections to the Schr\"odinger equation
\begin{align}
    i\partial_0\psi=&\hat{H}_{\text{0,NR,EM}}\psi-
    \frac{\ell}{2M}\sum_{n=1}^3\xi_n (Mc)^{3-n}\hat{k}_\mathcal{A}^n\psi,\label{eqn:DefSchrodFromKG}
\end{align}
where the dimensionless numbers $\xi_n$ parameterise the result in accordance with Eq. \eqref{eqn:xis}. Thus, the nonrelativistic evolution exactly equals Eq. \eqref{eqn:DefCanSchroEl}.

\section{Phenomenological implications in the infrared}\label{sec:PhenoInf}

Summarising the above, we have derived nonrelativistic quantum dynamics in two complementary ways, as presented in Eqs. \eqref{eqn:DefCanSchroEl} and \eqref{eqn:DefSchrodFromKG}, converging to the nonrelativistic Hamiltonian operator
\newline
\begin{align} 
    \tikzmarkin{first}(0.6,1.9)(-0.6,-1.8)
\hat{H}_{\ell,\rm{NR},\rm{EM}}\psi=&
    \left[-\frac{h^{ij}}{2M}\left(\nabla_i-ieA_i^{e}-iMA_i^{g}\right)\left(\nabla_j-ieA_j^{e}-iMA_j^{g}\right)+e\phi_e+M\phi_{\rm g}\right.\nonumber\\
    &\left.-\frac{\ell}{2M}\sum_{n=1}^3\xi_n(Mc)^{3-n}\left[-h^{ij}\left(\nabla_i-ieA_i^{e}-iM\mathcal{A}_i\right)\left(\nabla_j-ieA_j^{e}-iM\mathcal{A}_j\right)\right]^{\frac{n}{2}}\right]\psi.
\tikzmarkend{first}\label{eqn:HLNRFinal}
\end{align}
\newline
Equation \eqref{eqn:HLNRFinal} constitutes the central result of this paper. We emphasise that this Hamiltonian encompasses the nonrelativistic dynamics of all analytic MDRs up to order $\ell.$
 
The focus of this section is on analysing this unified Schrödinger equation as to report which kinds of models serve as a reliable foundation for low-energy quantum gravity phenomenology. Indeed, there are some conclusions that can be drawn from Eq. \eqref{eqn:HLNRFinal} at the general level, without specifying the considered MDR. Amongst these are the results depicted in Fig. \ref{fig:UVIR}.
\begin{itemize}
    \item Being dependent on fractional powers of the second-order differential operator $\hat{k}_{\mathcal{A}}^2,$ the corrections to the Hamiltonian are generally nonanalytic and therefore nonlocal. The appearance of such nonanalyticities is not a novelty in linearly deformed nonrelativistic quantum dynamics and generally expected at the nonperturbative level. For instance, they have been encountered in the context of the linear and quadratic GUP (see \eg \cite{Ali:2010yn}).
    \item As in the nonrelativistic regime by construction $\langle h^{ij}\hat{k}_i\hat{k}_j\rangle\ll (Mc)^2$, those MDRs which lead to linear corrections in $\hat{k}_\mathcal{A},$ in Eq. \eqref{mdr1} the ones linear in $k,$ \ie deformations of the form $E^2k,$ $MEk$ or $M^2k$, have the largest effects. Therefore, they are displayed in frames in Fig. \ref{fig:UVIR}. The corresponding Hamiltonians have the form
\begin{equation}
    \hat{H}_{\ell,\text{NR,EM}}\simeq \hat{H}_{0,\text{NR,EM}}-\frac{\xi_1\ell M c^2}{2} \hat{k}_\mathcal{A},\label{HlNR1}
\end{equation}
    such that the ensuing quantum-gravity contribution to the dynamics has the advantage of being amplified by a factor of $c^2.$ However, this also implies that the corrections become large in the far IR. If, for simplicity, $\mathcal{A}_i=0,$ this is the case once $\langle\hat{k}\rangle \lesssim \xi_1 \ell (Mc)^2.$ In fact, this property is shared with the underlying MDR which at small momenta becomes
    \begin{equation}
        M^2c^4\simeq E^2-k^2c^2-\xi_1\ell M^2c^4 k\simeq E^2-\xi_1\ell M^2c^4 k.\label{eqn:MDRa210}
    \end{equation}
    This finding entails an entirely new string of IR-phenomenology which has already been emphasised in \cite{Carmona:2000gd,Amelino-Camelia:2008aez}.
    \item Corrections quadratic in $\hat{k}_\mathcal{A},$ stemming from the models $Mk^2,~Ek^2$ (corresponding to the bicrossproduct basis of the $\kappa$-Poincar\'e algebra \cite{Kowalski-Glikman:2002iba}), $~M^2E,~ME^2$ and $E^3$, lead to a modification akin to the kinetic term. The underlying Hamiltonian is of the form
\begin{equation}
    \hat{H}_{\ell,\text{NR,EM}}\simeq \hat{H}_{0,\text{NR,EM}}-\frac{\xi_2\ell M c^2}{2} \hat{k}_\mathcal{A}^2.\label{HlNR2}
\end{equation}
In comparison to the ordinary kinetic term, here the mass is amended by a term comparing $M$ to $c/\ell$ (\ie the Planck mass), and the particle is minimally coupled to the vector potential $\mathcal{A}^i.$ If $\mathcal{A}^i$ and $N^i$ vanish, this comes down to a modification of the inertial mass, thus leading to a violation of the equivalence principle. We will come back to this point in section \ref{sec:WEP}.
\item Corrections proportional to $k_\mathcal{A}^3,$ \ie MDRs of the form $k^3$ occurring in LIV-models \cite{Kostelecky:2008ts} and a lesser-known basis of the $\kappa$-Poincar\'e algebra \cite{Girelli:2006fw,Amelino-Camelia:2014rga}, are special in their own right. Indeed, for isotropic backgrounds ($\mathcal{A}_i=0$), they exactly correspond to the kind of correction induced by the linear GUP \cite{Ali:2009zq,Ali:2011fa}, \ie
\begin{equation}
    \hat{H}_{\ell,\text{NR,EM}}\simeq \hat{H}_{0,\text{NR,EM}}-\frac{\xi_3\ell}{2M}\hat{k}_\mathcal{A}^3.
\end{equation}
In other words, this class of models predicts the customary minimal-length contribution at lowest order. Remarkably, those are the models which are least amenable to nonrelativistic tests, roughly going like $v^2/c^2$ (with the speed of the object at hand $v$) with respect to those which are linear in $\hat{k}_{\mathcal{A}}.$ Therefore, they are the only ones on the violet side of Fig. \ref{fig:UVIR}.
\item The relative velocity between the reference system and the deformation vector $\mathcal{A}^i$ is a central characteristic of the deformation-induced contributions to the evolution. This is where LIV and DSR diverge in interpretation. While the application of DSR avoids the presence of a preferred frame such that the vector $\mathcal{A}_i$ is probably unobservable,\footnote{While vectors of this kind do arise in DSR, they transform in a special way under the deformed Lorentz transformations such that their orientation with respect to the observer does not change \cite{Amelino-Camelia:2022dsj}. If the preferred direction, for instance, is to the observer's right, it will be so in every reference frame.} LIV requires a preferred frame. For the moment, we concentrate on the latter. Considering planet earth as the reference as, for example, required to describe table-top experiments, barring cosmic conspiracies, this dependency implies universal seasonal changes to most of the existing observables. Indeed, it is well-known not only that the earth moves relative to the rest frame of the cosmic microwave background at a speed of $\sim 10^{-3}c$ but also that this speed undergoes changes of order $1$ along the earth's trajectory. Hence, if the deformation vector $d^\mu$ does not vary along the way exactly such that this effect is continuously cancelled out, amounting to the above-mentioned cosmic conspiracy, we can generally estimate that
\begin{equation}
    \mathcal{A}\equiv \sqrt{h^{ij}\mathcal{A}_i\mathcal{A}_j}\geq 10^{-3}c\label{eqn:MinVel}
\end{equation}
for considerable time-spans during the year. While this speed is nonrelativistic, it results comparably large with respect to the dynamics of ordinary nonrelativistic systems. Therefore, in many relevant cases for phenomenology we can approximate $\hat{k}_{\mathcal{A}}\simeq M\mathcal{A}.$  This is particularly interesting because, then, in terms of the kinetic term $\hat{H}_{\text{kin}}=\hat{k}^2/2M,$ the correction terms $\hat{H}_n$ assume the form
\begin{equation}
    \hat{H}_n=-\xi_n\ell(Mc)^{3-n}\frac{\hat{k}_{\mathcal{A}}^n}{\hat{k}^2}\hat{H}_{\text{kin}}\simeq -\xi_n\ell M^3 c^{3-n}\frac{\mathcal{A}^n}{\hat{k}^2}\hat{H}_{\text{kin}}.
\end{equation}
In comparison to the ordinary kinetic term their contribution becomes larger, the smaller the velocities of the involved particles relative to the given reference system are. This introduces a strong amplifier for macroscopic quantum objects which generally move more slowly. Indeed, on the classical level, we obtain
\begin{equation}
    H_n\geq -\xi_n 10^{-3n}\ell Mc \left(\frac{c}{v}\right)^2\hat{H}_{\text{kin}},
\end{equation}
with the velocity $v.$ If $10^{-3n}\ell Mc\sim 10^{-19-3n}$ as, for example for the proton (whose mass is generally compared to the Planck mass also for macroscopic objects, \cf Appendix \ref{app:Mac}), this factor may soon be overcome by precise experimentation involving small velocities.
\end{itemize}
In a nutshell, with Eq. \eqref{eqn:HLNRFinal} we have a unifying description of MDRs in the nonrelativistic regime at our disposal, and ranked the underlying MDRs by amenability to nonrelativistic experiments leading up to Fig. \ref{fig:UVIR}. Furthermore, we have already found a connection to minimal-length models. In the subsequent section, we deepen this connection by translating the corrections into the language of GUPs. 

\section{Relation to minimal-length models}\label{sec:GUP}

As expressed in Eq. \eqref{eqn:HLNRFinal}, the nonrelativistic limit of MDRs harbours quantum-gravity inspired deformations of the single-particle Hamiltonian. The kinematics, in turn, is undeformed. Schematically, this structure can be outlined as
\begin{align}
    [\hat{x}^i,\hat{k}_j]=i\delta^i_j,&&\hat{H}=\frac{\hat{k}^2}{2M}+V(\hat x)+\delta H(\hat{k}).\label{eqn:GUPWaveNumber}
\end{align}
In the context of minimal-length theories, this way of displaying the corrections is known as the wave-number representation \cite{Bosso:2023sxr}, owing to the fact that the wave number $\hat{k}$ is \emph{defined} to be the conjugate variable to the position. The momentum $\hat{p}$, in turn, is understood to be a function of the wave number. The customary starting point then lies in deforming the Heisenberg algebra (between $\hat{x}^i$ and $\hat{p}_j$), while assuming the Hamiltonian to appear undeformed in the phase-space coordinates underlying that algebra, \ie
\begin{align}
    [\hat{x}^i,\hat{p}_j]=if^i_j(\hat p),&&\hat{H}=\frac{\hat{p}^2}{2M}+V(\hat{x}),\label{eqn:GUPMomentum}
\end{align}
for some tensor-valued function $f^i_j.$ As long as the coordinates $\hat{x}^i$ commute,\footnote{The commutator of the coordinates is exactly specified by a Jacobi identity given a function $f^i_j,$ see \cite{Wagner:2021bqz} for more details.} the schemes in Eqs. \eqref{eqn:GUPWaveNumber} and \eqref{eqn:GUPMomentum} are related by the noncanonical transformation $\hat{k}_i\rightarrow\hat{p}_i,$ $\hat{x}^i\rightarrow \hat{x}^i.$ Therefore, as long as the Jacobian $J^i_j=\partial \hat{p}_j/\partial \hat{k}_i=f^i_j$, is nondegenerate, models of the kind given in Eq. \eqref{eqn:GUPWaveNumber} can be reexpressed as a minimal-length deformation of the Heisenberg algebra with commuting coordinates. 

Note that this transformation essentially corresponds to a relabelling of momenta. However, if the vector $\mathcal{A}_i$ is time-dependent, the operator $\hat{p}_i$ acquires an intrinsic time-dependence. This implies that the Hamiltonian $\hat{H}$ cannot be the generator of time evolution of $\hat{p}_i$ just because $\mathcal{A}_i$ is an external field, whose dynamics is not governed by $\hat{H}$. Instead, $\hat{p}_i$ satisfies the Heisenberg equation $\dot{\hat{p}}_i=-i[\hat{p}_i,\hat{H}]+\partial \hat{p}_i/\partial t.$
Indeed, in this case $\hat{H},$ similarly to the Hamiltonian describing motion on time-dependent backgrounds, \eg in cosmology, is not a conserved quantity. For the phenomenological setting, which we are dealing with in the present paper, \ie in local experiments and for short times, we consider the vector $\mathcal{A}_i$ to be static at first approximation.\footnote{By analogy with the discussion in Sec. \ref{sec:TimeDepBack}, the time-dependence of $\mathcal{A}_i$ is not allowed to be strong. Otherwise, the single-particle approximation would be inappropriate.}

For the deformations we obtained in Eqs. \eqref{eqn:DefCanSchroEl} and \eqref{eqn:DefSchrodFromKG} (ignoring for the moment the unrelated issues with time-dependent backgrounds), and to first order in $\ell,$ we can express the momentum operator as
\begin{equation}
    \hat{p}_i=\hat{k}_i-\frac{\ell}{4}\left\{\hat{k}_i-2M\mathcal{A}_i+\frac{M^2}{2}\left\{\mathcal{A}^2,\frac{\hat{k}_i}{\hat{k}^2}\right\},\frac{\xi_1(Mc)^2}{\hat{k}_\mathcal{A}}+\xi_2 Mc+\xi_3\hat{k}_\mathcal{A}\right\},\label{eqn:GUPMoml}
\end{equation}
where $\mathcal{A}^2=h^{ij}\mathcal{A}_i\mathcal{A}_j,$ to obtain a structure of the kind given in Eq. \eqref{eqn:GUPMomentum}. This implies that the canonical commutation relations are deformed as
\begin{align}
    [\hat{x}^i,\hat{p}_j]=&i\delta^i_j\left[1-\frac{\ell}{4}\left\{1+\frac{M^2}{2}\left\{\mathcal{A}^2,\hat{p}^{-2}\right\},\frac{\xi_1(Mc)^2}{\hat{p}_\mathcal{A}}+\xi_2 Mc+\xi_3\hat{p}_\mathcal{A}\right\}\right]\nonumber\\
    &+\frac{i\ell M^2}{4}\left\{\left\{\frac{\hat{p}^i\hat{p}_j}{\hat{p}^3},\mathcal{A}^2\right\},\frac{\xi_1(Mc)^2}{\hat{p}_\mathcal{A}}+\xi_2 Mc+\xi_3\hat{p}_\mathcal{A}\right\}\nonumber\\
    &-\frac{i\ell}{4}\left\{\hat{p}_\mathcal{A}^i\left(-\frac{\xi_1(Mc)^2}{\hat{p}^3_\mathcal{A}}+\frac{\xi_3}{\hat{p}_\mathcal{A}}\right),\hat{p}_j-2M\mathcal{A}_j+\frac{M^2}{2}\left\{\mathcal{A}^2,\frac{\hat{p}_j}{\hat{p}^2}\right\}\right\},\label{eqn:MDRDefHeis}
\end{align}
where we introduced the operator $\hat{p}_i^\mathcal{A}=\hat{p}_i-M\mathcal{A}_i.$ As Eq. \eqref{eqn:MDRDefHeis} is, admittedly, involved, it is instructive to revert to the isotropic case $\mathcal{A}_i=0.$ As a result, the momentum operator $\hat{p}_i$ can be expressed as
\begin{equation}
    \hat{p}_i|_{\mathcal{A}_i=0}=\hat{k}_i\left[1-\frac{\ell}{2}
    \left(\frac{\xi_1(Mc)^2}{\hat{k}}+\xi_2 Mc+\xi_3\hat{k}\right)\right].
\end{equation}
By analogy with Eq. \eqref{eqn:MDRDefHeis}, the Heisenberg algebra is deformed as
\begin{equation}
    [\hat{x}^i,\hat{p}_j]=i\left[1-\frac{\ell}{2}\left(\frac{\xi_1(Mc)^2}{\hat{p}}+\xi_2 Mc+\xi_3\hat{p}\right)\right]\delta^i_j-\frac{i\ell}{2}\left(\frac{\xi_1(Mc)^2}{\hat{p}^3}+\frac{\xi_3}{\hat{p}}\right)\hat{p}^i\hat{p}_j.\label{eqn:MDRDefHeisA0}
\end{equation}
In the context of the literature on minimal-length models the deformed Heisenberg algebra given in Eq. \eqref{eqn:MDRDefHeis} as well as its simplified version (Eq. \eqref{eqn:MDRDefHeisA0}) are unusual in several aspects. A number of comments are in order.
\begin{itemize}
    \item The appearance of $\mathcal{A}_i$ in Eq. \eqref{eqn:MDRDefHeis} indicates a degree of anisotropy in the model induced by the external vector $d^\mu$. Anisotropic minimal-length models, while having been considered recently \cite{Gomes:2022hva,Gomes:2022tcd}, are rather uncommon. As highlighted in in the reasoning leading up to Eq. \eqref{eqn:MinVel}, in LIV the vector $\mathcal{A}_i$ is expected to undergo seasonal changes (and therefore be time-dependent) by a magnitude of the order $10^{-3}c$ which amounts to a large velocity in comparison to conventional nonrelativistic dynamics. Yet, if the model is complemented by deformed Lorentz-transformations under which it is symmetric, the nonrelativistic model is expected to inherit a symmetry under a deformation of rotations so that the anisotropy may only be apparent, see \eg \cite{Amelino-Camelia:2023rkg}. 
    \item For nonvanishing $\mathcal{A}_i,$ but also for isotropic models with nonvanishing $\xi_1,$ the deformed Heisenberg algebra contains inverse powers of the momentum operator. Considering, for simplicity, $\mathcal{A}_i=\xi_2=\xi_3=0,$ we have
    \begin{equation}
        [\hat{x}^i,\hat{p}_j]=i\left[1-\frac{\xi_1\ell(Mc)^2}{2\hat{p}}\right]\delta^i_j+\frac{i\ell\xi_1(Mc)^2}{2\hat{p}}\frac{\hat{p}^i\hat{p}_j}{p^2}.\label{eqn:MDRDefHeisAxi2xi30}
    \end{equation}
    Both correction terms in this case go like $\hat{p}^{-1}.$ This implies that the models do not reduce to ordinary quantum mechanics in the small-momentum limit, \ie $\lim_{\hat{p}_i\to 0}[\hat{x}^i,\hat{p}_j]\neq i\delta^i_j.$ As the commutator of positions and momenta exactly equals the Jacobi matrix, this implies that the noncanonical transformation $\hat{x}^i\rightarrow\hat{x}^i,$ $\hat{k}_i\rightarrow\hat{p}_i$ is degenerate at this point. In other words, our perturbative expansion in $\ell$ breaks down because the correction to the nonrelativistic Hamiltonian, Eq. \eqref{eqn:HLNRFinal}, becomes larger than the kinetic term -- note that this is not an issue at the relativistic level where the dispersion relation additionally contains the rest mass (\cf Eq. \eqref{eqn:MDRa210}). In minimal-length model building the appearance of inverse powers of $\hat{p}$ is usually avoided due to the technical and conceptual issues nontrivial IR effects entail. However, these problems do not indicate that the model is wrong \emph{per se}. As mentioned in the preceding section, this effect may just be the result of UV/IR-mixing (see \cite{Minwalla:1999px,Szabo:2001kg}) in low-energy quantum gravity phenomenology (\cf Sec. of \cite{Amelino-Camelia:2008aez} and \cite{Carmona:2000gd}).
    \item Customary minimal-length models correspond to the sector $\mathcal{A}_i=\xi_1=\xi_2=0$ of the full parameter space, thus deriving from MDRs of the kind
    \begin{equation}
        M^2c^4=E^2-k^2c^2-\xi_3\ell c^2 k^3,
    \end{equation}
    and can be turned into the deformed Heisenberg algebra
    \begin{equation}
        [\hat{x}^i,\hat{p}_j]=i\left[1-\frac{\xi_3\ell\hat{p}}{2}\right]\delta^i_j-\frac{i\ell\xi_3\hat{p}}{2}\frac{\hat{p}^i\hat{p}_j}{p^2},\label{eqn:MDRDefHeisAxi1xi20}
    \end{equation}
    which is equivalent to the first-order approximation of the linear and quadratic GUP \cite{Ali:2009zq} with model parameter $\alpha=2\xi_3$. Similarly, the quadratic GUP, involving commutative coordinates, at second order in $\ell$ assumes the form
    \begin{equation}
        [\hat{x}^i,\hat{p}_j]\simeq i\left[(1+\beta\ell^2\hat{p}^2)\delta^i_j+2\beta\hat{p}^i\hat{p}_j\right],
    \end{equation}
    with the dimensionless quadratic GUP-parameter $\beta.$ These results have two consequences. 
    
    First, it is noteworthy that all existing constraints on minimal-length models can be readily employed to constrain the parameter $\xi_3=a_{0,0,3}$ and vice versa. The reciprocal transfer of information between minimal-length models and MDRs imposes rather stringent limitations on the linear GUP-parameter $\alpha$ (and, by extension, on its quadratic counterpart $\beta$). Notably, time-delay measurements, for instance of the gamma-ray burst GRB 190114C \cite{MAGIC:2020egb}, the blazar PKS 2155-304 \cite{HESS:2011aa}, and the Crab pulsar \cite{MAGIC:2017vah} yield constraints on the order of magnitude
    \begin{equation}
    \alpha < \mathcal{O}(1) \quad \text{and} \quad \beta < \mathcal{O}(10^{16}).
    \end{equation}
    Consequently, these findings establish stringent bounds on the linear GUP reaching up to the Planck scale, while also improving upon previous constraints for the quadratic GUP by approximately 17 orders of magnitude.\footnote{Previously, scanning tunnelling microscope experiments \cite{Pikovski:2011zk, Das:2008kaa} had provided the most stringent constraints of $\beta<10^{33}$ in this context (see \cite{Bosso:2023aht} for an up-to-date collection of bounds).}
    
    Second, there is a clear sense in which the MDRs given in Eq. \eqref{eqn:mdr0} are more general than conventional minimal-length models. 
    \item As the relation to minimal-length models is instantiated by a redefinition of the momentum phase-space variable, the underlying coordinates are forced to commute. This continues to be the case on the nonperturbative level. General isotropic\footnote{Here we only consider isotropic models for clarity. The underlying reasoning could be straightforwardly generalised to anisotropic ones.} minimal-length models encompass noncommutative geometries.
    Hence, in a certain sense, minimal-length models are also more general than MDRs.
\end{itemize}
The link between minimal-length models and MDRs offers notable advantages by enabling the transfer of constraints from the latter to the former. However, it is worth emphasising that nonrelativistic considerations can also contribute to enhancing the constraints on MDRs. This aspect is exemplified in the subsequent subsection.

\section{Applications}

A phenomenological model has to be considered incomplete if it fails to provide reasonably testable predictions regarding observable phenomena. In the following analysis, we delve into the implications arising from deformations proposed in Eq. \eqref{eqn:HLNRFinal}. Initially, we focus on the simplest systems where scalar potentials vanish, and the background metric remains time-independent. To facilitate comparison with laboratory-based tests, we will also present the results in the context of flat-space and Newtonian limits. Additionally, we dedicate considerable attention to the equivalence principle.

\subsection{Vanishing scalar potentials -- Landau levels}\label{sec:Land}

First, to facilitate a detailed analysis of the deformed dynamics described by Eq. \eqref{eqn:HLNRFinal}, we make a number of simplifying assumptions. Specifically, we assume that the electric and gravitational potentials ($\phi$ and $\phi_{\rm g}$) vanish, and the Lorentz-violation is aligned along the global time-direction ($\mathcal{A}_i=0$). With these assumptions in place, the corrections in the dynamics depend solely on the undeformed Hamiltonian. In other words, we obtain
\begin{equation}
    \hat{H}_{\ell,\text{NR,EM}}=\hat{H}_{\text{0,NR,EM}}-\frac{\ell}{2M}\sum_{n=1}^{3}\xi_n(Mc)^{3-n}(2M\hat{H}_{\text{0,NR,EM}})^{n/2}.
\end{equation}
It is reasonable to expect that this property continues to hold at the nonperturbative level.\footnote{Here, the term "nonperturbative" refers to expansions in both $\ell$ and $1/c.$} Consequently, stationary states remain unmodified. Since $\hat{H}_{\ell,\text{NR,EM}}=\hat{H}_{\ell,\text{NR,EM}}(\hat{H}_{0,\text{NR,EM}})$, it is evident that the eigenstates of $\hat{H}_{0,\text{NR,EM}}$ are identical to the eigenstates of $\hat{H}_{\ell,\text{NR,EM}}$. Yet, their eigenenergies are modified as
\begin{equation}
    E_{\ell,k_i}=E_{0,k_i}-\frac{\ell}{2M}\sum_{n=1}^{3}\xi_n(Mc)^{3-n}(2ME_{0,k_i})^{n/2}=E_{0,k_i}-\frac{\ell}{2M}\sum_{n=1}^{3}\xi_n c^{3-n}M^{3-\frac{n}{2}}(2E_{0,k_i})^{n/2},\label{shift-energy}
\end{equation}
with the eigenvalue of the unperturbed Hamiltonian $E_{0,k_i},$ which depends on some possibly discrete variable $k_i$, the eigenvalue of the momentum operator. Note that this modification of the energy can, in general, not be absorbed into a redefinition of the time parameter because this redefinition would be dependent on the mass of the particle at hand. Once we consider particles of different masses, the correction reappears. Curiously, an energy dependence of physical ``constants'' may be possible in the rainbow-gravity approach \cite{Magueijo:2002xx}.

As an illustrative example, we consider the configuration of a particle with charge $-|e|$ in a spacetime endowed with a static electric or magnetic field. Specifically, we can examine the scenario of a cosmic string coupled to an axial magnetic field \cite{Vilenkin:1981zs, Furtado:1994np}. In cylindrical coordinates, the line element is given by
\begin{equation}
\D s^2=-\D t^2+\D\rho^2+\alpha^2\rho^2\D\phi^2+\D z^2\, ,
\end{equation}
where $\alpha\leq 1$ measures the angular deficit of this string, which defines a spacetime with conical structure. The vector potential that furnishes a magnetic field $\Vec{B}=\Vec{\nabla}\times \Vec{A}$ with constant norm $h_{ij}B^iB^j=B^2$, defined just in the $z$-direction, has only one non-zero component $A_{\phi}(\rho)=B\rho/(2\alpha)$. The energy levels of this system are characterised by the discrete and continuous variables $k_i=(n,l,k_z),$ where $n\in {\mathbb N}$ represents the radial, $l\in {\mathbb Z}$ the angular and $k_z\in {\mathbb R}$ the $z$-component of the momentum. For a detailed derivation of the solution and the corresponding energy levels in the Minkowski case, we refer the reader to \cite{ciftja2020detailed}, and for the cosmic string correction to \cite{Furtado:1994np} from which we recover the energy levels
\begin{equation}
    E_{0,(n,l,k_z)}=\frac{\hbar \omega_B}{2\alpha}\left(2n+\frac{|l|}{\alpha}-\frac{l}{\alpha}+1\right)+\frac{\hbar^2k_z^2}{2M}\, ,
\end{equation}
where $\omega_B=|e|B/Mc$ is the cyclotron frequency of the configuration. Plugging this expression into Eq. \eqref{shift-energy} implies the shifted energy levels
\begin{equation}
    E_{\ell,(n,l,k_z))}=E_{0,(n,l,k_z)}-\frac{\ell}{2M}\sum_{n'=1}^{3}\xi_{n'} c^{3-n'}M^{3-\frac{n'}{2}}(2E_{0,(n,l,k_z)})^{n'/2}.\label{shift-energy2}
\end{equation}

The variable $k_i$ is of broader utility in the enumeration of stationary states, allowing for the expression of a general state at $t=0$, denoted as $\psi(0)$, as follows
\begin{equation}
    \psi(0)=\SumInt_{k_i} a_{k_i}\psi_{k_i} (0),
\end{equation}
where the sum (integral) is over the discrete (continuous) enumerator $k_i,$ and the coefficients $a_{k_i}$ are chosen such that $\SumInt_{k_i} |a_{k_i}|^2=1.$ Then, the corresponding time evolved state (at $t=\Delta t$) becomes
\begin{equation}
    \psi (\Delta t)=\SumInt_{k_i} e^{iE_{\ell,k_i}\Delta t}a_\mathcal{P}\psi_{k_i} (0).
\end{equation}
As a result, these corrections introduce additional relative phases between stationary states during time evolution, which makes them suitable for exploration through interferometric table-top experiments. In this vein, the duration of the time evolution $\Delta t$ can be used as amplifier for the effect. We hope to report back on this subject in the future.

\subsection{Flat spacetime and table-top experiments}\label{sec:FlatLab}

In flat spacetime, there exists a foliation where $N=1$, $N^i=0$, and $h_{ij}=\delta_{ij}$. Therefore, we straightforwardly obtain the deformed flat-space Schr\"odinger equation
\begin{align}
    i\partial_0\psi=\left[-\frac{\Delta_0}{2M}+e\phi_e-\frac{\ell}{2M}\sum_{n=1}^{3}\xi_n(Mc)^{3-n}(-\Delta_\mathcal{A})^{n/2}\right]\psi,\label{eqn:SchrodFlat}
\end{align}
with the $\mathcal{A}$-deformed gauge-covariant flat-space Laplacian $\Delta_{0,\mathcal{A}}$. In Cartesian coordinates, this differential operator can be expressed as
\begin{equation}
    \Delta_{0,\mathcal{A}}=\delta^{ij}\left(\partial_i-ieA^e_i-iM\mathcal{A}_i\right)\left(\partial_j-ieA^e_i-iM\mathcal{A}_j\right).
\end{equation}
The ordinary gauge-covariant Laplacian $\Delta_0$ can be derived from it as $\Delta_0=\Delta_{0,\mathcal{A}}|_{\mathcal{A}_i=0}$.
Furthermore, in flat space, we can translate the deformed Schr\"odinger equation into the momentum representation
\begin{equation}
    i\partial_0\tilde{\psi}=\left[-\frac{k^2}{2M}+e\phi_e(i\dot{\partial}^i)-\frac{\ell}{2M}\sum_{n=1}^{3}\xi_n(Mc)^{3-n}k_{\mathcal{A}}^n\right]\tilde{\psi},\label{eqn:SchrodFlatMom}
\end{equation}
with the momentum-derivative $\dot{\partial}^i=\partial/\partial k_i.$ The shift to the momentum-representation removes the intricacies of fractional powers of differential operators. The ensuing corrections are universal in the same way that GUP-corrections are \cite{Das:2008kaa}. This allows for a plethora of different applications to table-top experiments. 

Among others, in the context of minimal-length models, for example, there have been constraints from implementations of Planck-scale corrections to the harmonic oscillator \cite{Chang:2001kn,Dadic:2002qn,Bosso:2021koi,Ali:2011fa}, the hydrogen atom \cite{Benczik:2005bh,Slawny:2007pya,Chang:2011jj,AntonacciOakes:2013qvs,Brau:1999uv}, the scanning tunnelling microscope \cite{Pikovski:2011zk,Das:2008kaa}, the $\mu$ anomalous magnetic moment \cite{Das:2011tq,Das:2008kaa}, the Landau levels \cite{Ali:2011fa,Das:2008kaa,Das:2009hs} (\cf Sec. \ref{sec:Land}), the lamb shift \cite{Das:2008kaa,Ali:2011fa}, $^{87}$Rb interferometry \cite{Gao:2016fmk,Khodadi:2018kqp}, the Kratzer potential \cite{Bouaziz:2013ora}, stimulated emission \cite{Twagirayezu:2020chx}, quantum noise \cite{Girdhar:2020kfl} and the Tsirel'son bound \cite{Bosso:2022ogb}. A wide range of examples, including those mentioned above and many others, can be explored within the broader framework of MDR-induced deformations. 

Here, we will content ourselves with corrections to energy eigenvalues of the isotropic harmonic oscillator under the assumption that $\mathcal{A}_i=0,$ thus outlining a procedure for general analyses into central potentials for isotropic Lorentz-violation. At the unperturbed level, and in spherical coordinates, the isotropic harmonic oscillator allows for the stationary states \cite{Messiah:1961js}
\begin{equation}
    \tilde{\psi}_{nlm}=R_{nl}(k)Y_l^m(\theta,\phi),\label{eqn:HarmOscWave}
\end{equation}
with the spherical harmonics $Y_l^m$ and the radial component
\begin{equation}
    R_{nl}(k)=N_{nl}k^lL_{\frac{n-l}{2}}^{l+\frac{1}{2}}\left(\frac{k^2}{M\omega}\right)e^{-\frac{k^2}{2M\omega}}.
\end{equation}
Here, we introduced the normalisation factor
\begin{equation}
    N_{nl}=2\sqrt{\frac{2^{\frac{n+l}{2}}(M\omega)^{-\frac{3}{2}-l}(\frac{n-l}{2})!}{\sqrt{\pi}(n+l+1)!!}}.
\end{equation}
Moreover, $\omega$ and $L_n^a(x)$ stand for the oscillation frequency and the generalised Laguerre polynomials, respectively. The integers $(n,l,m),$ where $n\geq 0,$ $l=\mod(n,2),\mod(n,2)+2,\dots n,$ $-l\leq m\leq l,$ denote the quantum numbers, in terms of which the unperturbed eigenenergies read
\begin{align}
    E_{n}=\omega \left(n+\frac{3}{2}\right).
\end{align}
Applying time-independent perturbation theory \cite{Sakurai:2011zz}, the corrections to the eigenvalues can be obtained as
\begin{equation}
    \Delta E_{nl}=-\frac{\ell}{2M}\sum_{o=1}^3\xi_o(Mc)^{3-o}\langle\psi_{nlm}|\hat{k}^o\psi_{nlm}\rangle.\label{eqn:HarmOscInt}
\end{equation}
Thus, we have to calculate the expectation value of powers of $\hat{k},$ which effectively comes down to evaluating a momentum-space integral in spherical coordinates. The angular integration is trivial, such that the expectation value simplifies to
\begin{equation}
    \langle\hat{k}^o\rangle=N_{nl}^2\int_0^\infty\D k  k^{o+2}R_{nl}^2(k).\label{eqn:HarmOscInt2}
\end{equation}
Indeed, this kind of expression for corrections to the spectrum of the Hamiltonian is generic for every central potential. In this specific case, the integral is over polynomials multiplying a Gaussian akin to moments of a normal distribution. Thus, introducing the dimensionless $\bar{k}=k/\sqrt{M\omega},$ we can solve the integral in all generality with the help of an analogue to a source $J$ such that
\begin{align}
    \langle\hat{k}^o\rangle=&(M\omega)^{\frac{3+2l+o}{2}}N_{nl}^2\lim_{J\to0}\frac{\partial^{o+2l+2}}{\partial J^{o+2l+2}}\left[L_{\frac{n-l}{2}}^{\frac{1}{2}+l}\left(\frac{\partial^2}{\partial J^2}\right)\right]^2\int_0^\infty\D \bar k  e^{-\bar{k}^2+J \bar{k}},\\
    =&(M\omega)^{\frac{o}{2}}\frac{2^{\frac{2+n+l}{2}}\left(\frac{n-l}{2}\right)!}{(n+l+1)!!} 
    \lim_{J\to0}\frac{\partial^{o+2l+2}}{\partial J^{o+2l+2}}\left[L_{\frac{n-l}{2}}^{\frac{1}{2}+l}\left(\frac{\partial^2}{\partial J^2}\right)\right]^2e^{\frac{J^2}{4}}\left[1+\text{Erf}\left(\frac{J}{2}\right)\right],\label{eqn:HarmOscInt3}
\end{align}
with the error function Erf$(x).$ The generalised Laguerre polynomials have the closed-form expression
\begin{equation}
    L^{\alpha}_n(x)=\sum_{n'=0}^{n}(-1)^{n'}{\alpha+n\choose n-n'}\frac{x^{n'}}{n'!},
\end{equation}
with the binomial coefficient ${m\choose n}.$ Furthermore, derivatives of the generating function can be obtained in general, yielding
\begin{equation}
    \lim_{J\to0}\frac{\partial^o}{\partial J^o}e^{\frac{J^2}{4}}\left[1+\text{Erf}\left(\frac{J}{2}\right)\right]=\Gamma\left(\frac{o+1}{2}\right).
\end{equation}
As as result, we can express the expectation value of powers of the momentum as
\begin{equation}
    \langle\hat{k}^o\rangle=(M\omega)^{\frac{o}{2}}\frac{2^{\frac{2+n+l}{2}}\left(\frac{n-l}{2}\right)!}{(n+l+1)!!}\sum_{n',n''=0}^{\frac{n-l}{2}}(-1)^{n'+n''}{\frac{n+l+1}{2}\choose \frac{n-l-n'}{2}}{\frac{n+l+1}{2}\choose \frac{n-l-2n''}{2}}\frac{\Gamma\left[\frac{3+o+2(l+n'+n'')}{2}\right]}{n'!n''!}.\label{eqn:error}
 \end{equation}
Note that $n-l$ is an even number such that the sum is well-defined. For the MDRs associated with $\xi_2,$ this expression simplifies significantly. In fact, it is possible to provide a closed-form expression even in the case of nonvanishing, but constant $\mathcal{A}_i$ such that
\begin{equation}
    \Delta E_{n}|_{\xi_1=\xi_3=0}=-\frac{\xi_2}{2}Mc\ell\left[M\mathcal{A}^2+\omega\left(n+\frac{3}{2}\right)\right]=-\frac{\xi_2}{2}Mc\ell\left[M\mathcal{A}^2+E_n\right].
    \end{equation}
Here, the corrections introduced by nonvanishing $\mathcal{A}_i$ are constant, and can therefore be absorbed into a redefinition of the zero-point energy.

The harmonic oscillator holds significance beyond its pedagogical value. In recent years, considerable research has focused on investigating the potential of macroscopic harmonic oscillators as a means to examine minimal-length deformations \cite{Pikovski:2011zk,Bawaj:2014cda,Bushev:2019zvw}. Notably, in \cite{Bawaj:2014cda}, bounds on the quadratic GUP-parameter $\beta$ were derived through such investigations. It is important to note that these analyses did not incorporate the necessary rescaling of the minimal length for compounds provided in Eq. \eqref{macr-h}. Taking into account this rescaling, the ensuing constraint becomes
\begin{equation}
    \beta<\mathcal{O}(10^{48}),\label{eqn:HarmOscBoundBeta}
\end{equation}
bearing testament to the elusiveness of minimal-length deformations in the nonrelativistic regime. In order to shed light on the potential implications of quantum gravity phenomenology at low energies, we aim to establish a preliminary estimate of how the obtained results could be translated into constraints on the various MDR-parameters $\xi_n$. To accomplish this, we begin by deriving an approximate bound on the linear GUP-parameter $\alpha$, which, as discussed in Section \ref{sec:GUP}, essentially corresponds to $\xi_3$. Subsequently, we proceed to approximate the resulting constraints on $\xi_1$ and $\xi_2$. We stress that these considerations cannot replace a full phenomenological derivation of the effect.

On the level of order-of-magnitude estimates, the linear and the quadratic GUP imply perturbative corrections $\Delta_{\text{GUP}} O$ to the predicted value of any observable $O$ which go like
\begin{equation}
    (\Delta O)_{\text{GUP}}\sim O\alpha\ell\bar{k}\quad\text{and}\quad(\Delta O)_{\text{GUP}}\sim O\beta\left(\ell\bar{k}\right)^2,
\end{equation}
respectively, with some momentum scale $\bar{k}.$ If this observable has been measured to a relative precision $\delta O=\Delta O/O,$ and no deviation from the case $\ell=0$ has been detected, the model parameters can be constrained as
\begin{equation}
    \alpha\sim \frac{\delta O}{\ell\bar{k}}\quad\text{and}\quad\beta\sim \frac{\delta O}{\left(\ell\bar{k}\right)^2}.\label{eqn:PrecGUP}
\end{equation}
Above, we have provided an upper bound to the quadratic GUP of the form $\beta\leq\beta_0.$ By virtue of Eq. \eqref{eqn:PrecGUP}, such a bound can be roughly translated to the linear GUP as
\begin{equation}
    \alpha\leq \sqrt{\beta_0\delta O}.
\end{equation}
In \cite{Bawaj:2014cda}, the observable in question is given by the frequency $\omega$ which is measured to a relative precision of $\delta\omega\sim 10^{-7},$ thus leading to a bound 
\begin{equation}
    \alpha\leq 10^{21}.
\end{equation}
Bounds on the other MDR-parameters can be obtained considering that we can estimate $\langle p^\sigma\rangle\sim (Mv)^\sigma,$ with the velocity $v\sim \bar{A}\omega,$ and the maximal amplitude $\bar{A}.$ From this consideration, we infer that bounds on $\xi_3$ may be roughly translatable to the other MDRs by the recipe
\begin{equation}
    \alpha\sim\xi_3 \sim \frac{c}{v}\xi_2\sim\left(\frac{c}{v}\right)^2\xi_1.
\end{equation}
As expected, we see an enhancement of the effect for other MDRs. For instance, in \cite{Bawaj:2014cda}, the considered oscillator moves at a velocity $v\sim 10^{-11}c.$ Thus, our preliminary considerations indicate that we may obtain bounds of the orders of magnitude
\begin{align}
    \xi_1\leq \mathcal{O}(10^{-1}),&&\xi_2\leq \mathcal{O}(10^{10}),&&\xi_3=\mathcal{O}(10^{21}).
\end{align}
Thus, even though the input constraint in Eq. \eqref{eqn:HarmOscBoundBeta} appears to be far below the Planck scale, it could actually rule out those MDRs associated with the parameter $\xi_1.$ We find a clear indication that nonrelativistic experiments are more than competitive in some areas of quantum gravity phenomenology.

\subsection{Corrections to Newtonian gravity and the weak equivalence principle}\label{sec:WEP}

Newton's law of gravity emerges as a limiting case of general relativity when we examine the solution to Einstein's field equations that describes the exterior of a massive object, the Schwarzschild metric
\begin{equation}
    \D s^2=-F(r)\D t^2+\frac{\D r^2}{F(r)}+r^2\D\Omega^2,
\end{equation}
with the metric the two-sphere $\D\Omega^2,$ and where $F=1+2\phi_N/c^2.$ Here, we introduced the Newtonian gravitational potential
\begin{equation}
    \phi_N=-\frac{G\mathcal{M}}{r},
\end{equation}
where $\mathcal{M}$ denotes the mass of the central object. In this background the shift-vector vanishes and the lapse function equals $N=\sqrt{F}.$ Quantum-gravity corrected, nonrelativistic dynamics on this background is therefore based on the Schr\"odinger equation
\begin{equation}
    i\partial_0\psi=\left[-\frac{\Delta_0}{2M}+M\phi_\text{N}+e\phi_e-\frac{\ell}{2M}\sum_{n=1}^{3}\xi_n(Mc)^{3-n}(-\Delta_{0,\mathcal{A}})^{n/2}\right]\psi.\label{DefNewton}
\end{equation}
In the following, we want to study the effect of the corrections on the equivalence principle.\footnote{That the weak equivalence principle could be a good tested for MDRs was already pointed out in \cite{Amelino-Camelia:2009wvc}.} In doing so, we assume those corrections to be isotropic, \ie $\mathcal{A}_i=0,$ and neglect the influence of the electromagnetic interaction ($A_i^e=\phi_e=0$). As is customary at short distances from the surface of a large, massive body, we approximate the gravitational potential as $\phi=gz,$ with the gravitational acceleration $g,$ and where we introduced the coordinates $x^i=(x,y,z).$ From Eq. \eqref{DefNewton}, we can then read off the Hamiltonian
\begin{equation}
    \hat{H}_{\ell,\text{NR},\text{EM}}=\frac{\hat{k}^2}{2 M}+Mgz-\frac{\ell}{2M}\sum_{n=1}^3\xi_n(Mc)^{3-n}\hat{k}_{\mathcal{A}}^n,
\end{equation}
Assuming that, generally, $\mathcal{A}\simeq \text{const}$ along the duration of experiments and $M\mathcal{A}\gg \hat{k}$ (applying to all considered correlation functions), the ensuing Heisenberg equations indicate a deformation of the expected acceleration $a=\langle{\ddot{\hat{x}}^z}\rangle$ 
\begin{equation}
    M_{\text{I}}a=-Mg\equiv M_{\text{g}}g,
\end{equation}
with the gravitational mass $M_{\rm g}=M$ and the effective inertial mass\footnote{At the level of perturbation theory (in $\ell$), it is irrelevant whether we attribute the corrections to the gravitational mass or the inertial mass.}
\begin{equation}
    M_{\text{I}}\simeq M\left[1+\frac{\ell Mc}{2}\left(\xi_1c\frac{\mathcal{A}^2-\mathcal{A}_z^2}{\mathcal{A}^3}+2\xi_2+3\xi_3\frac{\mathcal{A}^2+\mathcal{A}_z^2}{\mathcal{A}c}\right)\right].
\end{equation}
Thus, the weak equivalence principle, one of the most precisely tested predictions in all of physics, is violated.

Considering free fall of two distinct massive bodies $A$ and $B,$ this violation of the weak equivalence principle can be summarised in the E\"otv\"os parameter
\begin{equation}
    \eta(A,B)=2\left\langle\frac{\frac{M_{\text{g},A}}{M_{\text{I},A}}-\frac{M_{\text{g},B}}{M_{\text{I},B}}}{\frac{M_{\text{g},A}}{M_{\text{I},A}}+\frac{M_{\text{g},B}}{M_{\text{I},B}}}\right\rangle\simeq \frac{c\ell}{2}(M_B-M_A)\left(\xi_1c\frac{\mathcal{A}^2-\mathcal{A}_z^2}{\mathcal{A}^3}+2\xi_2+3\xi_3\frac{\mathcal{A}^2+\mathcal{A}_z^2}{\mathcal{A}c}\right).\label{eqn:eotvosSP}
\end{equation}
Note that the corrections associated with $\xi_2$ are independent of the value of $\mathcal{A}$ altogether. We further assume $\mathcal{A}_z^2-\mathcal{A}^2\sim \mathcal{A}_z^2+\mathcal{A}^2\sim \mathcal{A}^2,$ in favour of which we can invoke the same argument as for the lower bound to $\mathcal{A}$ entertained in Sec. \ref{sec:PhenoInf}. Even more so, tests of the equivalence principle have been performed not only at manifold different times, but also on differing locations of the planet for all of which the coordinate $z,$ pointing towards the interior of the planet, corresponds to distinct directions. Then, as an order-of-magnitude estimate the E\"otv\"os parameter generically has the lower bound
\begin{equation}
    \eta(A,B)\geq c\ell(M_B-M_A)\left(10^3\xi_1+\xi_2+ 10^{-3}\xi_3\right).\label{eqn:eotvosSP}
\end{equation}
As for common tests of the equivalence principle, the studied objects are nonrelativistic in the lab frame. 

So far, the E\"otv\"os parameter has been constrained most effectively by the MICROSCOPE collaboration \cite{Touboul:2017grn}. From their latest data \cite{MICROSCOPE:2022doy}, it is clear that
\begin{equation}
    |\eta|<10^{-14}.
\end{equation}
Eventually, its sensitivity is expected to improve this bound by a further order of magnitude \cite{Touboul:2017grn,MICROSCOPE:2022doy}. 

Before comparing to our result, however, we have to take into account that the Schr\"odinger equation, Eq. \eqref{DefNewton}, applies to single elementary particles, not to composites. The details of this result can be found in Appendix \ref{app:Mac}; analogous results for the GUP have been reported in \cite{Quesne:2009vc,Tkachuk:2012gyq,Amelino-Camelia:2013fxa} (for a more recent treatment see \cite{Bosso:2023aht}). The MICROSCOPE collaboration, however, works with macroscopic objects. In line with Eq. \eqref{macr-h}, we can include the deterioration with increasing number of constituents $\mathcal{N}$ by rescaling the minimal length as $\ell\to \ell/\mathcal{N}$ (with the number of elementary constituents $\mathcal{N}$) such that the E\"otv\"os parameter satisfies
\begin{equation}
    \eta(A,B)\geq \frac{c\ell}{\mathcal{N}}(M_B-M_A)\left(10^3\xi_1+\xi_2+ 10^{-3}\xi_3\right).
\end{equation}
The number of elementary constituents per gram of matter, in turn, is proportional to Avogadro's number $\mathcal{N}\sim 10^{23}.$ Furthermore, the MICROSCOPE experiment involves mass differences of order $1 \text{kg}.$ Altogether, we obtain the rough constraints
\begin{align}
    |\xi_1|\leq \mathcal{O}(10^1)&&|\xi_2|\leq \mathcal{O}(10^4), && |\xi_3|\leq \mathcal{O}(10^{7}).
\end{align}
Thus, MDRs associated with $\xi_1$ are bound up almost to the Planck scale, while those associated with $\xi_2,$ containing the bicrossproduct basis of DSR are constrained to four orders of magnitude below. We stress that this latter result is independent of the value of $\mathcal{A},$ and, therefore, also applies to theories without preferred frame. Furthermore the MDR leading to corrections going with $\xi_3,$ conventionally associated with LIV, has been bounded to seven orders of magnitude below Planckian precision. For comparison, the bound on GUP-like MDRs, purporting $\mathcal{A}=0$ are $21$ orders of magnitude away from the Planck scale, \ie rather weak.

The suppression with increasing number of particles can be circumvented by microscopic determination of $\eta.$ This is achieved for the comparison of free fall for single atoms \cite{Dimopoulos:2006nk,Zhou:2011kah,Kovachy:2015jak} (see \cite{Tino:2020dsl} for a recent review), where $\mathcal{N}\sim 1.$ In this context, the best results achieved thus far constrain $|\eta|\leq \mathcal{O}(10^{-12})$ \cite{Asenbaum:2020era}. We thus obtain the bound
\begin{align}
    |\xi_1|\leq \mathcal{O}(10^5)&&|\xi_2|\leq \mathcal{O}(10^8), && |\xi_3|\leq \mathcal{O}(10^{11}),
\end{align}
which is generally worse than the one stemming from macroscopic objects. However, this field may yield possibilities for improvement in the future \cite{Dimopoulos:2006nk,Zhou:2011kah,Kovachy:2015jak}. We find that the change in the number of fundamental constituents can be overcome by the decrease in the difference of masses (in latter case taken between the masses of single $\text{Rb}^{85}$ and $\text{Rb}^{87}$ isotopes). As explained in Appendix \ref{app:Mac} this can be understood from the point that the quantity $M/\mathcal{N}$ roughly provides the mass of the constituents dominated by protons.

\section{Conclusion}  

Modified dispersion relations take a prominent place amongst the candidate effects of quantum gravity at low energy. Up until now, they have mainly been investigated at scales beyond the LHC center-of-mass energy. However, both the precision of low-energy measurements as well as the control over quantum states have seen significant improvements in recent times. Assuming that this progress continues, tests of quantum gravity in the nonrelativistic quantum regime will become increasingly feasible. In the language of \cite{Amelino-Camelia:2008aez}, the energy is just one amongst many amplifiers, and it may soon be overcome by precision, statistics, time of evolution, number of coherent particles, or combinations thereof.

Therefore, in this paper we have performed a $1/c$-expansion of MDR-inspired dynamics at first order in the Planck length. In so doing, we have remained agnostic on the specific model by solving for all analytic MDRs at that level. In particular, we have followed two complementary paths to arrive at the $1/c$-expansion. First, we have derived the nonrelativistic limit of MDRs at the classical level to subsequently apply canonical quantisation. Second, inspired by the underlying MDR, we have provided a deformed Klein-Gordon equation, describing the single-particle sector of bosonic quantum field theory, and taken its nonrelativistic limit.

We have found, that the results agree for backgrounds with non-evolving induced geometries on the hypersurfaces. While single-particle quantum dynamics on time-dependent backgrounds is subtle for several reasons, these problems are not specific to MDRs and can, for the most part, be neglected. 

We have further compared our results with the pertinent approach to quantum gravity phenomenology in the nonrelativistic regime, minimal-length models. As had already been pointed out in \cite{Hossenfelder:2007fy}, GUPs with commutative geometry are just a special case of MDRs. In fact, minimal-length models are associated with those MDRs which have the weakest effects at nonrelativistic energies, being suppressed by powers of $1/c.$ To corroborate this observation, we have applied high-energy constraints from astrophysical time-delay measurements of MDRs to constrain the dimensionless minimal-length model parameters ($\alpha$ for the linear GUP \cite{Ali:2009zq}, $\beta$ for the quadratic GUP \cite{Kempf:1994su}, order one amounts to the Planck scale) as $\alpha < \mathcal{O}(1)$ and $\beta < \mathcal{O}(10^{16}).$ Hence, while the linear GUP is ruled out up to the Planck scale, the constraints on the quadratic model are improved by 17 orders of magnitude. 

Finally, we have applied the formalism to several instructive and/or phenomenologically relevant configurations. In particular, we have considered the Landau levels in static spacetimes, where corrections have been obtained in all generality. Furthermore, we have dealt with flat backgrounds which allow for a rich phenomenology regarding table-top experiments. In particular, we have calculated corrections to the spectrum of the harmonic oscillator and estimated the impact of GUP-bounds on other MDRs. The results suggest that some MDRs may be constrained up to the Planck scale by nonrelativistic experiments. To provide a further example, we have applied the formalism to the Schwarzschild geometry, and obtained the corrections to  the Schr\"odinger equation in a Newtonian gravitational field. Consequently, we encountered violations of the weak equivalence principle which allowed for the bounds on several MDRs reaching up to one order of magnitude below the Planck scale, while those MDRs generally associated with the bicrossproduct basis of DSR \cite{Kowalski-Glikman:2002iba} are constrained to energy scales of $10^{15}$GeV. These findings demonstrate that nonrelativistic experiments can be competitive in quantum gravity phenomenology even beyond the Bose-Marletto-Vedral proposal \cite{Bose:2017nin,Marletto:2017kzi}.

On a different note, a number of popular MDRs are indistinguishable in the far UV due to the fact that $E\simeq kc.$ For instance, the bicrossproduct basis of the $\kappa$-Poincar\'e algebra leads to corrections of the form $Ek^2/c$ \cite{Kowalski-Glikman:2002iba}, while models of LIV \cite{Kostelecky:2008ts} as well as a lesser-known basis of the $\kappa$-Poincar\'e algebra \cite{Girelli:2006fw,Amelino-Camelia:2014rga} predict contributions going like $k^3.$ At high energies, both deformations lead to essentially equal effects; diverging predictions only emerge on the level of relative locality \cite{Amelino-Camelia:2011lvm} and deformed translations \cite{Rosati:2015pga}, however the analysis of this scenario is still in its initial stages \cite{Bolmont:2022yad}. Yet, in the IR the energy satisfies approximately $E\simeq M c^2,$ clearly distinguishing the former from the latter kinds of MDR. Sizeable violations of the weak equivalence principle, for example, only follow from the MDR of the type $Ek^2.$

To put it in a nutshell, MDRs have a rich phenomenology in the IR which remains to be explored. The GUP, in turn, whose nonrelativistic effects have been studied extensively over the last 30 years, is particularly amenable to observations in the UV, \emph{not} the IR. Those MDRs, which imply strong effects in the nonrelativistic regime, merit further investigation. We hope to report back on this issue in the future. 


\section*{Acknowledgements}
I. P. L. was partially supported by the National Council for Scientific and Technological Development - CNPq grant 306414/2020-1 and by the grant 3197/2021, Para\'iba State Research Foundation (FAPESQ). The authors would like to acknowledge networking support by the COST Action QGMM (CA18108), supported by COST (European Cooperation in Science and Technology). V. B. B. was partially supported by the National Council for Scientific and Technological Development - CNPq grant 307211/2020-7. G.V.S was supported by the National Council for Scientific and Technological Development - CNPq grant 140335/2022-6.

\appendix

\section{Deformed macroscopic dynamics}\label{app:Mac}

The dynamics implied by Eq. \eqref{eqn:HLNRFinal} applies solely at the microscopic level, \ie to elementary particles, and has to be amended to account for macroscopic objects. To elucidate this point, in this appendix we study its classical counterpart in the absence of external fields and curvature, and with aligning Lorentz-violation ($\mathcal{A}_i=0$). The resulting Hamiltonian reads
\begin{equation}
    H=\frac{1}{2M}\left(k^2-\ell\sum_{n=1}^3\xi_n (Mc)^{3-n}k^n\right),
\end{equation}
where $k^2=\delta^{ij}k_ik_j.$ Applying the Legendre transform, we obtain the Lagrangian
\begin{equation}
    L=\frac{Mv^2}{2}\left[1+\ell Mc \left(\xi_1 \frac{c}{v}\xi_2 Mc+\xi_3 Mv \right)\right],
\end{equation}
with the velocity $v$. In order to study the kinematics of macroscopic objects, we consider $\mathcal{N}$ particles of equal mass such that
\begin{equation}
    L=\sum_{I=1}^\mathcal{N}\frac{Mv_I^2}{2}\left[1+\ell Mc \left(\xi_1 \frac{c}{v_I}\xi_2 Mc+\xi_3 Mv_I \right)\right]
\end{equation}
In this context, the velocity of a single particle can be expressed as a sum of the macroscopic center-of-mass velocity $v_{\rm{mac}}$ and the relative velocity of the particle. Assume, for simplicity, the relative motion to be negligible. We are, thus, describing a macroscopic solid in motion. As a result, on the macroscopic level the Lagrangian becomes
\begin{equation}
    L=\sum_{I=1}^\mathcal{N}\frac{Mv_{\rm{mac}}^2}{2}\left[1+\ell Mc \left(\xi_1 \frac{c}{v}\xi_2 Mc+\xi_3 Mv \right)\right]=\frac{M_{\rm{mac}}v_{\rm{mac}}^2}{2}\left[1+\frac{\ell M_{\rm{mac}}c}{\mathcal{N}}\left(\xi_1\frac{c}{v_{\rm{mac}}}+\xi_2+\xi_3\frac{v_{\rm{mac}}}{c}\right)\right],\label{macr-h}
\end{equation}
where, in the last step, we introduced the mass of the macroscopic object $M_{\rm{mac}}=\mathcal{N}M$. Hence, expressed in terms of the relevant variables for macroscopic objects, the quantum-gravity corrections deteriorate with the number of fundamental constituents, thereby avoiding a possible soccer-ball problem \cite{Hossenfelder:2007fy,Amelino-Camelia:2011dwc,Hossenfelder:2014ifa,Amelino-Camelia:2014gga}.\footnote{Note that a resolution of the soccer-ball problem of this kind creates a new ``inverse'' soccer-ball problem, for details see \cite{Bosso:2023aht}.} As regards phenomenology, we effectively have to shift the relevant length scale as $\ell\to\ell/\mathcal{N}.$ As we generally compare this diluted minimal-length scale to the macroscopic mass, this comes down to comparing the mass of the fundamental constituents to the Planck scale. As this mass is generally dominated by protons and the neutrons (both of which only encompass three elementary particles), it is generically the proton mass that is compared to the Planck mass. This fact which is compatible with other findings in the context of minimal-length models \cite{Quesne:2009vc,Tkachuk:2012gyq,Amelino-Camelia:2013fxa,Bosso:2023aht} and has to be accounted for, when considering macroscopic objects.

\providecommand{\href}[2]{#2}\begingroup\raggedright\endgroup

\end{document}